\documentclass[fleqn,usenatbib]{mnras}
\usepackage{float}  
\usepackage{times}
\usepackage{amsmath} 
\usepackage{graphicx}
\usepackage{comment}
\usepackage{xcolor}
\usepackage{booktabs}
\usepackage{multirow}
\usepackage[toc,page]{appendix}
\color[rgb]{0.0,0,0.0}
\usepackage{pifont} 
\usepackage{natbib}
\usepackage[switch]{lineno}
\usepackage{subcaption}
\usepackage{hyperref}
\usepackage{amssymb}
\usepackage{footmisc}
\usepackage{amsmath,amsfonts}
\usepackage{makecell}
\usepackage{subcaption}
\usepackage{tikz}
\usepackage{arydshln}
\usetikzlibrary{shapes.geometric, arrows, positioning}
\newif\ifAMStwofonts

\def\gtorder{\mathrel{\raise.3ex\hbox{$>$}\mkern-14mu
             \lower0.6ex\hbox{$\sim$}}}
\def\ltorder{\mathrel{\raise.3ex\hbox{$<$}\mkern-14mu
             \lower0.6ex\hbox{$\sim$}}}

\title{Towards sub-milliarcsecond astrometric precision using high-cadence seeing-limited imaging}
\author[Segev et al.]{
Noam Segev$^1$\footnotemark[1] ,
Eran O. Ofek$^1$,
Yossi Shvartzvald$^1$\footnotemark[2],
Krzysztof A. Rybicki$^1$,
Chung-Uk Lee$^2$\footnotemark[2],
\newauthor
Dong-Jin Kim$^2$\footnotemark[2],
Jennifer C. Yee$^3$\footnotemark[2],
Michael D. Albrow$^4$\footnotemark[2],
Sun-Ju Chung$^2$\footnotemark[2],
Andrew Gould$^{5}$\footnotemark[2],
\newauthor
Cheongho Han$^6$\footnotemark[2],
Kyu-Ha Hwang$^2$\footnotemark[2],
Youn Kil Jung$^{2,7}$\footnotemark[2],
In-Gu Shin$^3$\footnotemark[2],
Hongjing Yang$^{8,9}$\footnotemark[2],
\newauthor
Weicheng Zang$^3$\footnotemark[2],
Sang-Mok Cha$^{2,10}$\footnotemark[2],
Hyoun-Woo Kim$^2$\footnotemark[2],
Seung-Lee Kim$^2$\footnotemark[2],
Yoon-Hyun Ryu$^2$\footnotemark[2],
\newauthor
Dong-Joo Lee$^2$\footnotemark[2],
Yongseok Lee$^{2,10}$\footnotemark[2],
Byeong-Gon Park$^2$\footnotemark[2],
Richard W. Pogge$^{5,11}$\footnotemark[2]
\\
$^1$Department of Particle Physics and Astrophysics, Weizmann Institute of Science, Rehovot 7610001, Israel\\
$^2$Korea Astronomy and Space Science Institute, Daejeon 34055, Republic of Korea\\
$^3$Center for Astrophysics | Harvard \& Smithsonian, 60 Garden St., Cambridge, MA 02138, USA\\
$^4$University of Canterbury, School of Physical and Chemical Sciences, Private Bag 4800, Christchurch 8020, New Zealand\\
$^5$Department of Astronomy, Ohio State University, 140 W. 18th Ave., Columbus, OH 43210, USA\\
$^6$Department of Physics, Chungbuk National University, Cheongju 28644, Republic of Korea\\
$^7$National University of Science and Technology (UST), Daejeon 34113, Republic of Korea\\
$^8$School of Science, Westlake University, Hangzhou, Zhejiang 310030, China\\
$^{9}$Department of Astronomy, Tsinghua University, Beijing 100084, China\\
$^{10}$School of Space Research, Kyung Hee University, Yongin, Kyeonggi 17104, Republic of Korea\\
$^{11}$Center for Cosmology and AstroParticle Physics, Ohio State University, 191 West Woodruff Ave., Columbus, OH 43210, USA
}

\date{Accepted ?
      Received ?
      in original form ?}

\begin{document}
\maketitle

\begin{abstract}

The Earth's atmospheric turbulence degrades the precision of ground-based astrometry.
Here we discuss these limitations and propose that, with proper treatment of systematics and by leveraging the many epochs available from the Korean Microlensing Telescope Network (KMTNet), seeing-limited observations can reach sub-milliarcsecond precision.
Such observations may be instrumental for the detection of Galactic black holes via microlensing.
We present our methodology and pipeline for precise astrometric measurements using seeing-limited observations.
The method is a variant of Gaia's
Astrometric Global Iterative Solution (AGIS)
that include several detrending steps.
Tests on 6,500 images of the same field, obtained by KMTNet with typical seeing condition of 1 arcsecond and pixel scale of 0.4\,arcsecond, suggest that we can achieve, at the bright end (mag $\lesssim$17), per-epoch relative astrometric precision of $\mathord{\sim}$5\,mas and relative proper motion precision of 0.1-0.2\,mas\,yr$^{-1}$ over a baseline of approximately five years, using data from the Cerro Tololo Inter-American Observatory (CTIO) site. Time binning on 5--20~day cadences improves the bright-source precision to $\sim$2~mas per coordinate on astrometric microlensing-relevant timescales. 
The precision is estimated using bootstrap simulations and further validated by comparing results from two independent KMTNet telescopes.

\end{abstract}
\begin{keywords}
astrometry --- gravitational lensing: micro --- techniques: image processing --- stars: black holes --- atmospheric effects --- proper motions
\end{keywords}

\footnotetext[1]{E-mail: noam.segev@weizmann.ac.il}
\footnotetext[2]{The KMTNet Collaboration}

\section{Introduction}

For the past two millennia, precision astrometry provided a plethora of discoveries.
However, due to the Earth's turbulent atmosphere, ground-based astrometry is considerably less efficient (and even less cost-effective) compared with space-based astrometry.
The main problems of ground-based astrometry are:
(i) The starting point of the precision (i.e., given a single photon) is the seeing
rather than the diffraction limit;
(ii) Due to atmospheric scintillations, the photons arriving from the stars are spatially correlated,
and therefore the measurement precision decrease slower than $1/\sqrt{N_{\rm ph}}$, where $N_{\rm ph}$ is the number of photons.
Although adaptive optics observations and interferometric techniques (e.g., \citealt{genzel1997nature,Ghez+2005ApJ_StarOrbit_BH_GalCenter,gravity_interferometry_2010,dong2019_microlens_resolve_first}) can partially solve these problems, they are currently expensive and limited to a small field of view.
These two facts make it currently difficult for ground-based astrometry to compete with space-based astrometry.
However, even the leading space astrometry mission,  {\it Gaia} (\citealt{GAIA+2016_GAIA_mission}), has some limitations. Two such limitations are the capability of providing
astrometric measurements over short time scales ($\lesssim$ a few months), and the relative inefficiency of {\it Gaia} in crowded fields, such as the Galactic bulge.
Indeed, several science cases can benefit from precision astrometry measured on time scales
shorter than a few months, and/or targets in the Galactic bulge.
Examples include searching for astrometric microlensing due to compact objects (e.g., \citealt{Lu+2016_astrometricMicrolensing_KeckAO, Sahu+2017_astrometricMicrolensing_WD_mass_HST,rybicki_bh_ml_detection_2018,sahu2022isolated}),
searching for lensed quasars and measuring their time delays (e.g., \citealt{Springer+Ofek2021_TimeDelayI_FluxOnly,Springer+Ofek2021_TimeDelayII_FluxAstrometry}),
and searching for binary asteroids (e.g., \citealt{Segev+2023MNRAS_AstrometricDetection_BinaryAsteroids_GAIADR2}).

In this work, we develop and test a method for precision astrometry using high-cadence, ground-based observations of microlensing events towards the Galactic bulge.
Our main motivation for the development of this methodology is to find more examples of isolated stellar-mass black holes (BH),
and eventually to measure their mass function.
This is of great importance for our understanding of the core-collapse supernova explosion mechanism (e.g., \citealt{Heger+2003ApJ_HowMassiveStarsEndLife_Supernovae_CoreCollapse}) and the birth of BH binaries and gravitational wave events (e.g., \citealt{Abbott+2016_LIGO_FirstGW_BH_Merger}).
To date, most of the known stellar mass black holes are detected only in binary systems such as X-ray binaries (e.g., \citealt{BH_xray_binary_mcclintock2003black} and \citealt{xray_binary_ozel2010black}) and BH-BH (or BH-neutron star) mergers by Gravitational Waves (e.g., \citealt{BH_merger_abbott2016observation}). Three quiescent stellar-mass black holes were identified through their relatively wide-orbit stellar companions in {\it Gaia} observations \citep{elbadri_2023sun_bh_a,elbadri2023red_bh_b,panuzzo2024discovery_bh}.
To study the abundance of black holes in the Galaxy using binary systems, a complete understanding of the system's evolution is required.
Unfortunately, the evolution of such binary systems is complicated and likely includes physical processes that are not fully understood. Even with a complete understanding, binary-based methods would remain inherently limited, as they trace only those black holes that remain in observable binaries.

\cite{Blaes+Madau1993ApJ_AccretingISM_IsolatedNS_Xray} and \cite{Agol+2002_Xrays_IsolatedBlackHoles} suggested that isolated compact objects can be detected via X-ray emission that takes place due to Bondi-Hoyle-like accretion from the interstellar medium (ISM).
However, searches for such objects have not revealed good candidates so far (\citealt{Maoz+1997MNRAS_NewClass_EUV,chisholm2003stellar_no_detection_xray_bh} and \citealt{mereghetti2022xray_followup_bh_ogle_110462}).
Furthermore, even if such objects are identified, measuring their numbers and masses may prove extremely challenging due to the complexities of accretion physics.

The study of microlensing events is currently the most promising path for detecting isolated stellar-mass black holes.
To date, tens of thousands of gravitational microlensing events have been detected by the flux magnification using ground-based telescopes, which observe the Galactic Bulge at high cadence, e.g., OGLE \citep{ogle_mission_udalski1992optical}; MOA \citep{MOA_II_abe2008moa}; KMTNet \citep{kim2016kmtnet}. However, the microlensing light curve only rarely provides enough information for the lens mass to be inferred. 
There are two approaches to deal with this problem. The first is the statistical approach
in which, based on the statistical properties of the microlensing population
(e.g., the velocity distribution of lenses and sources) one can infer the mass distribution
of a sample of microlensing events.(e.g., \citealt{Gould2000_MeasuringRemnantMassFunction_Microlensing,mroz_wyrzykowski_2021measuring,mroz2021measuring,rybicki2024analysis}).
The second approach is breaking the degeneracy and measuring the lens mass.
Because, in the simplest case, there are two observables (the magnification and time scale),
and five unknowns (lens mass, source distance, lens distance, impact parameter, and relative tangential velocity between the source and the lens), the problem is degenerate.
For a Galactic microlensing event, the source distance can be estimated, and therefore, two 
observables are missing.
One such observable is the microlensing parallax \citep{refsdal1966possibility_parallax_satelite,ml_plx_gould1992extending}, and another is a measurement of the Einstein radius
in angular units.
The microlensing parallax can be measured for any event with sufficiently long time scales (e.g., \citealt{alcock1995first_first_microlens_parallax}) or simultaneously using long-baseline space-based observations (e.g., \citealt{dong2007first_microlens_parallax,zhu2015spitzer_microlens_parallax_mass}). Measuring the Einstein radius in angular units is, however, more difficult. One solution is to use the finite-source effect \citep{gould1994proper_finite_source_fs,witt_mao_1994_fs_effect,nemiroff1994_finite_source_fs}, but such events are extremely rare ($\mathord{\sim}10^{-3}$) for BH events. Resolving the lensed images is also difficult because the typical Einstein radius rarely exceeds a few milli-arcseconds (mas). However, this was already demonstrated using interferometric techniques and may become an industry in the coming years (e.g., \citealt{dong2019_microlens_resolve_first,cassan2022microlensing_mass_rotating_resolution, Wu+2024ApJ_Microlensing_Gracity_Resolved}).

Another approach, which is the main driver for our work, is to measure the position of the center-of-light of the lensed images over time. This motion deviates from pure proper motion combined with parallax, enabling the measurement of the Einstein radius in angular units~\citep{hog1995macho,miyamoto1995astrometry,walker1995microlensed,dominik2000astrometric}. Astrometric microlensing is observed as a relative effect, measured through the displacement of the source centroid with respect to the surrounding reference stars. Consequently, our analysis is formulated within a relative astrometric framework.
This approach was attempted using ground-based AO observations (e.g., \citealt{Lu+2016_astrometricMicrolensing_KeckAO}) and space-based observations.
So far, one isolated black hole has been found using {\it HST} and {\it OGLE} observations (\citealt{sahu2022isolated,lam2022isolated_ns_or_bh,lam2023reanalysis,mroz2022systematic}).

Detecting such events from the ground requires astrometric precision of $\lesssim1$\,mas, posing a significant observational challenge. However, there is one important advantage over other, more selective techniques. Ground-based astrometry is a by-product of the existing survey observations that are used for the discovery and monitoring of microlensing events. Hence, the search for astrometric microlensing signals can be applied to thousands of events rather than a few selected events with space-based or AO observations.

This balance may shift with the Nancy Grace Roman Space Telescope (\emph{Roman}), whose Bulge survey will deliver $\mathord{\sim}$15\,min cadence across six $\mathord{\sim}$70\,day seasons with per-epoch astrometry $\lesssim$1\,mas. Yet months-long gaps between seasons leave mid-duration ($\mathord{\sim}$60–70\,day) events under-sampled; ground surveys can supply dense coverage and early/late baselines. In that sense, our survey remains competitive in discovery volume and temporal coverage, and strongly complementary for characterization: it extends time baselines across \emph{Roman} seasons and provides an independent consistency check on \emph{Roman} astrometry. Because our program precedes \emph{Roman}, it is both a test bed and a complement.

To date, microlensing surveys (e.g., \citealt{ogle_mission_udalski1992optical, MOA_II_abe2008moa, kim2016kmtnet}) detect $\mathord{\sim}$3000 verified microlensing events per year. \cite{Gould2000_MeasuringRemnantMassFunction_Microlensing} estimates that $\mathord{\sim}$1\% of all microlensing events are due to stellar-mass black holes. Therefore, we expect that microlensing samples include numerous events in which the lens is an isolated BH, and the direct measurement of the Einstein radius may allow us to search for these events.
However, the detection of the astrometric microlensing signal requires typical precision which are of the order of milliarcseconds or below.
As far as we know, such precisions using seeing-limited ground-based astrometry were not reported in the past, although several attempts came close to the level of a few milliarcsecond precision (e.g., \citealt{monet1983ccd, Soszynski+2002AcA_OGLE_ProperMotion_LMC_SMC, Sumi+2004MNRAS_OGLE_ProperMotion_GalacticBuldge}).

In this paper, we present our method for precise astrometric measurements of seeing-limited ground-based data.
We demonstrate this method on data from the KMTNet telescopes (\citealt{kim2016kmtnet}).
This method can provide an astrometric precision of about 0.5\,mas on a large sample of events. This astrometric precision is likely enough to detect microlensing events with large Einstein radii. In companion papers, we apply this method to a large set of KMTNet data to measure proper motions and constrain the mass of microlensing events.

The astrometric microlensing signal typically extends over several times the Einstein crossing time $t_{\mathrm{E}}$, i.e., months to years depending on lens mass and relative kinematics. While the photometric magnification is confined to a period of order $t_{\mathrm{E}}$, the centroid shift remains detectable well before and after peak. For our purposes, the critical requirement is therefore not decade-long baselines, but sufficiently dense sampling across the times of maximum deflections as well as observations obtained a few Einstein crossing times before and after the maximum magnification, in order to constrain the proper motion.
The KMTNet survey provides exceptionally high cadence in the Galactic bulge ($\mathord{\sim}$10–200 epochs per field per day), enabling us to combine hundreds of measurements into effective astrometric epochs; in the ideal limit of uncorrelated noise, the centroid precision scales as $1/\sqrt{N}$. For example, binning 500 exposures (per-exposure precision $\sim$5\,mas) over a few days yields a nominal precision of $\approx 5/\sqrt{500}\simeq0.22$\,mas. In practice, residual systematics and short-timescale correlations limit this gain ($N \rightarrow N_{\rm eff} \le N$). Even so, over a typical months-long event, this strategy yields the equivalent of hundreds of high-precision astrometric epochs, hopefully sufficient to measure $\theta_{\mathrm{E}}$ with the accuracy required for our science goals.

We begin in \S\ref{sec:astrometricPrecision} by discussing the theoretical astrometric precision achievable under seeing-limited observations. The characteristics of KMTNet observations are described in \S\ref{sec:the_data}, followed by an outline of the astrometric catalogue extraction in \S\ref{sec:reduction}. The astrometric model, minimisation scheme, and the matrix form are presented in \S\ref{sec:astrometric_model}, \S\ref{sec:minimization}, and \S\ref{sec:matrix_form}, respectively. The systematic effects and detrending procedures are discussed in \S\ref{sec:detrending}. A step-by-step description of the whole algorithm appears in \S\ref{sec:step_by_step_algo}. Results from testing on real data are shown in \S\ref{sec:results}. Finally, \S\ref{sec:conclusion} summarises and discusses future directions.

\section{Theoretical astrometric precision of seeing-limited observations}
\label{sec:astrometricPrecision}

Before describing our astrometric reduction procedure, it is worthwhile to discuss the expected astrometric precision.
In the best-case scenario, the precision of astrometric measurements is bounded by the Cramér–Rao inequality; in the photon-limited regime, this bound reduces to the familiar Poisson limit, where the astrometric noise (per axis) is roughly $\approx\sigma_{\rm PSF}/(S/N)_{\rm meas}$.
Here $\sigma_{\rm PSF}$ is the $\sigma$-width of a Gaussian-like point spread function (PSF; i.e., $\sigma_{\rm PSF}\cong FWHM/2.35$), where the FWHM is the Full-Width at Half Maximum, $(S/N)_{\rm meas}$ is the $S/N$ for a measurement process: $S/N=N_{\rm ph}/\sqrt{(N_{\rm ph} + B)}$, $N_{\rm ph}$ is the number of photons from the source and $B$ is the variance of all the background contributions.
For example, for a source with a FWHM of approximately $1''$ and a signal-to-noise ratio ($S/N$) of 1000, the \textit{Poisson-limited precision} predicts an astrometric accuracy of 0.4\,mas. However, in reality, Earth's turbulent atmosphere causes changes in the photon's direction, and the spatial distribution of photons from a point source is highly correlated on timescales of tens of milliseconds. Furthermore, the phase scintilations have 
 an angular correlation scale of about ten arcseconds in visible light (i.e., the iso-planatic patch). For instance, in the Palomar Transient Factory (PTF, \citealt{PTF_paper_law2009palomar}), with $\sigma_{\rm PSF} = 1''$, $S/N \approx 1000$, and a 60-second exposure, the typical astrometric accuracy relative to GAIA is about 14 mas per axis (\citealt{Ofek2019_Astrometry_Code}), which is approximately 1.5 orders of magnitude worse than the \textit{Poisson-limited prediction}. This level of accuracy occurs because Poisson noise dominates only when the number of photons from the source, per atmospheric coherence time (roughly 20 ms), is less than one. In the $V$-band, this happens for sources fainter than a magnitude of approximately $20 + 5\,\log_{10}{D}$, where $D$ is the telescope diameter in meters.

State-of-the-art space telescopes, e.g., {\it HST} and {\it JWST}, deliver sub-milliarcsecond astrometry. For {\it HST}, bulge microlensing analyses achieve $\sim$0.2$-$0.3\,mas precision per epoch \citealt{sahu2022isolated,lam2022isolated_ns_or_bh}; for {\it JWST}, the {\it NIRCam} geometric-distortion solution has precision at the $\lesssim$0.2\,mas level \citep{griggio2023photometry}.

Furthermore, at some level of astrometric precision, we may expect that systematic errors will kick in.
There are many potential reasons for systematics, including inhomogeneity in pixel size, sampling effects, color terms, and more (e.g., \citealt{LSST_tree_rings_beamer2015study}).

However, because observations taken over a long period of time are expected to be uncorrelated, 
averaging multiple epochs can improve the astrometric precision by $1/\sqrt{N_{\rm ep}}$, 
where $N_{\rm ep}$ is the number of epochs. This was demonstrated, for example, using data from 
the Large Array Survey Telescope \citep{Ofek+2023PASP_LAST_Overview,Ofek+2023PASP_LAST_PipeplineI}. 
For some KMTNet microlensing events, $N_{ep} \mathord{\sim} 5000$, comparable to the number of epochs in the fields analyzed in this work, 
and therefore we may expect measurements with a precision on the order of 0.1\,mas. 
Since most KMTNet events are monitored in denser fields with $N_{ep} \mathord{\sim} 15000$, 
the achievable precision in those cases could be even better.
In practice, however, some cases may be limited by systematic errors.

\section{The Data}
\label{sec:the_data}
We use observations from the Korea Microlensing Telescope Network (KMTNet; \citealt{kim2016kmtnet}), a global array of three identical 1.6-meter telescopes located at the Cerro-Tololo Inter-American Observatory (CTIO) in Chile, the South African Astronomical Observatory (SAAO) in South Africa, and the Siding Spring Observatory (SSO) in Australia. Each telescope is equipped with a mosaic of four 9k\,$\times$\,9k CCDs, yielding a field of view of approximately $4~\mathrm{deg}^2$ and a pixel scale of $0.4~\mathrm{arcsec\,pixel^{-1}}$.
The typical atmospheric seeing conditions at these sites are between $0.6''$--$1.0''$ at CTIO, $0.8''$--$1.4''$ at SAAO, and $1.0''$--$2.0''$ at SSO \citep{kim2016kmtnet}.

KMTNet's microlensing survey targets the Galactic bulge, covering a total area of $\sim$97\,deg$^2$ divided into overlapping fields observed with a tiered cadence strategy. Central fields are monitored at up to 4\,hr$^{-1}$ by each telescope, with overlapping pairs reaching effective cadences of $\sim$8\,hr$^{-1}$. Outer fields are observed less frequently, with cadences as low as 0.2\,hr$^{-1}$ \citep{kim2018kmtnet}.

KMTNet observes in both I-band (most of the time) and V-band (every $\mathord{\sim}$10 epochs). Over the eight years of 2016–2023, each field has accumulated between a few thousand and over $10^5$ images, depending on its assigned cadence. The network collects approximately 50,000--75,000 science images annually during the bulge season. 

In this work, we use I-band images from the KMTNet BLG17K0103 field, obtained with the CTIO observatory, to evaluate the pipeline performance and characterize systematics. Additionally, I-band images from the KMTNet BLG15M0306 field, taken with both the CTIO and SAAO observatories, are used to test the consistency between different sites. These two fields were chosen because they are relatively diluted and provide a cadence that is well-suited for astrometric analysis and pipeline development. Over the 2016–2023 baseline, BLG17K0103 contains about 7,000 images from CTIO, while the denser microlensing fields typically accumulate closer to 15,000 images. However, despite their higher cadence, the more crowded fields are less favourable for precise astrometric analysis and calibration of the pipeline, due to blending and stronger systematics.

For the development and testing of the astrometric pipeline, we extracted image stamps of $300\mathord{\times}300$ pixels (corresponding to $2^{\prime}\mathord{\times}2^{\prime}$). This choice was made to reduce computational runtime and because the point-spread function, although asymmetric, remains effectively constant over this spatial scale.

Table \ref{tab:ObservationsSum} summarizes the data we used in this work.

\begin{table*}
\caption{
Summary of the observations over the KMTNet fields used in this work. 
The number of epochs refers to the total number of observations, with the number of epochs used in the astrometric analysis shown in parentheses (after filtering for airmass and removal of poor-quality images). 
}
\label{tab:ObservationsSum}
\centering
\begin{tabular}{lllllll}
\hline
KMTNet field  & Total \# of epochs & \# CTIO epochs (used) & \# SAAO epochs (used) & Epochs range & RA [hms] & Dec [dms] \\[0.1cm]
\hline 
\noalign{\vskip 0.1cm}
BLG17K0103    & 18336 & 7671 (6557) & 5201 (-- ) & Feb-2016\,--\,May-2023 & 17:43:56.54& -32:52:15.42\\
BLG15M0306    & 18527 & 7737 (6283) & 5313 (4173) & Feb-2016\,--\,May-2023 & 17:40:04.17& -26:02:36.60 \\
\hline
\end{tabular}
\end{table*}

\section{Astrometric Catalogue Extraction}
\label{sec:reduction}

The pipelines begin with images that have been bias-subtracted and flat-field corrected. 
As part of this work, we are using tools from the MATLAB Astronomy \& Astrophysics Package\footnote{\url{https://github.com/EranOfek/AstroPack}}(\citealt{Ofek2014_MAAT, Soumagnac+Ofek2018_catsHTM, Ofek2019_Astrometry_Code, Ofek+2023PASP_LAST_PipeplineI}).
The first step is to estimate the image background\footnote{Using {\tt imProc.background.background}}. 
Instead of performing source detection in each image, we rely on the reference catalogue employed by KMTNet, which is constructed from a combination of the OGLE-III star catalogue \citep{szymanski2011optical_OGLE_III_catalog} and the DECam Plane Survey catalogue \citep{schlafly2018decam_catalog}.  This composite catalogue reaches a limiting magnitude of approximately 22.

The next step is to perform PSF photometry. 
The PSF is constructed from bright, isolated stars in the field, typically using several tens of stars to ensure a well-sampled empirical PSF. 
For each source, its position is measured using the weighted first moment, and it is shifted to the pixel origin using Whittaker-Shannon interpolation. Then, we normalize the cutout of each source and set the PSF as the median value in each pixel.
To avoid introducing artifacts caused by Whittaker-Shannon interpolation on a finite-size stamp, and artifacts caused by nearby stars, we employ a hybrid model to describe the PSF. 
The hybrid model is divided into two regimes, determined by the radius $\bar{r}$ from the center of the PSF. 
For the inner part ($r < \bar{r}$), the PSF is represented by the empirical PSF, while for the outer part ($r \mathord{\geq} \bar{r}$), we use a fit to an analytical model. The optimal transition radius, $\bar{r}$, varies with observing conditions and is typically close to the FWHM.
We adopt the Multivariate $t$-distribution (see \S\ref{sec:mtd}) as the analytic PSF model because it can account for PSF asymmetries. 

We construct a single PSF for each field because our pipeline operates on a small field of view (FOV) of about $2$\,arcmin. 
In such a small FOV, spatial variations in the PSF across the field are small.

We adopt a single, spatially invariant PSF per field because each cutout spans $\mathord{\sim}2'$ on a side, only $\mathord{\sim}10^{-4}$ of KMTNet’s $4\,\mathrm{deg}^2$ frame, over which the spatial PSF gradients (set mainly by smoothly varying optical aberrations) are small. This assumption breaks down for cutouts that span a significant fraction of the telescope’s field of view, where a spatially varying PSF model becomes necessary.

\begin{figure}\centering
\includegraphics[width=1\linewidth]{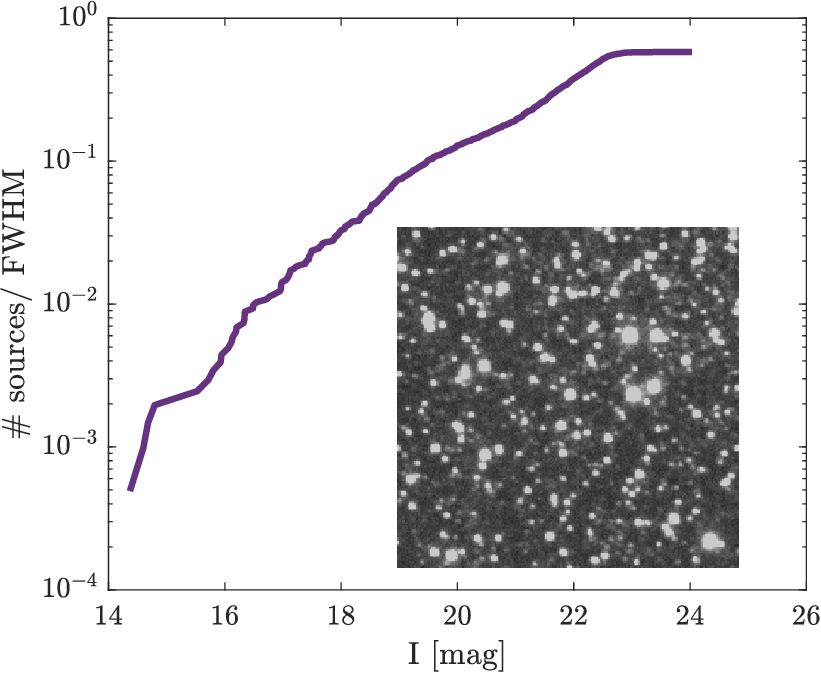}
\caption{Source density for KMTNet BLG17K0103 field. The density is calculated based on the KMTNet catalogue, and the units are the number of sources per PSF's FWHM area. For the calculation, we use a typical 1D FWHM of 3\,pix. We present an example of the $200\times200$\,pixels image cutout of the BLG17K0103 in the right-bottom corner.}
\label{fig:sources_density_fwhm}
\end{figure}

Given the crowded nature of the fields, as demonstrated in Figure \ref{fig:sources_density_fwhm}, and the significant effects of blending, we use a multi-iteration fit-and-subtract approach. 
Sources are divided into ten bins based on their estimated magnitudes. 
We start by fitting the brightest stars and subtracting them from the image, then proceed to fit stars in the next magnitude bin.

For PSF fitting, the $\chi^{2}$ between the observations and the model is minimized. 
The fit for each source includes three free parameters: $x$, $y$, and flux, while the background is fixed and measured only in the first iteration. 
It is worth noting that the photometry obtained from this process is not as accurate as that derived from image subtraction.

We note that our astrometric solution is done relative to an arbitrary reference frame (i.e., not the International Coordinates Reference System; ICRS).
Furthermore, this reference frame may have position-dependent proper motion relative to the ICRS, and any parallax measurements in this reference frame are relative, not absolute (i.e., $\pi = 0$ may correspond to a finite distance).

After extracting the catalogue we align the catalogue to a fiducial epoch\footnote{For the initial alignment, we use {\tt imProc.trans.fitPattern} to fit a discrete affine transformation (step size = $0.2$\,pix) from each of the catalogues to the fiducial catalogue (e.g., earliest catalogue).}.
Finally, we match\footnote{{\tt Using imProc.match.matchedReturnCat}} the aligned catalogues and construct a matrix of observations per epoch (row) for each source (column). We set the matching radius to $r=2$\,pix.

\section{Astrometric model}
\label{sec:astrometric_model}
Here, we describe the astrometric model and its uncertainties.
The model includes proper motion, parallax, Differential Chromatic Refraction (DCR), and instrumental distortions. 

All astrometric quantities in this work are defined relative to a fiducial reference frame constructed from the ensemble of stars in each field. This formulation follows the fact that our measurements, and those relevant for astrometric microlensing, are differential by construction, i.e., they describe relative displacements of sources with respect to the local reference frame rather than their absolute positions on the sky.

We first list the parameters relevant to the astrometric model. These are divided into two categories: measured quantities and fitted unknowns.

\textbf{Measured parameters:}
\begin{itemize}
    \item $i$ is the source index.
    \item $j$ is the epoch index.
    \item $t_j$ is the epoch.
    \item $x_i(t_j)$, $y_i(t_j)$ is the positions of source $i$ at epoch $t_j$.
    \item $\text{pa}_j$ is the parallactic angle
    \item $z_j$ is the zenith angle, and $\text{sec}(z_j)$ is the airmass.
    \item $\bar{C}\equiv V-I - median(V-I)$, where $V,I$ are the magnitude in the V and I bands, taken from the KMTNet catalogue.
    \item $\text{ha}_j,\text{alt}_j$ the hour angle and altitude of the target field.
\end{itemize} 

\textbf{Fitted unknown parameters:}\begin{itemize}
    \item $x_i^0, y_i^0, \mu_{x,i}, \mu_{y,i}$ are the source reference position and proper motion in x- and y-axis.
    \item $\pi_i$ is the parallax measure.
    \item $a_1^j, a_2^j ,a_3^j$ are the epochs' affine transformation parameters for the x-axis.
    \item $a_4^j, a_5^j ,a_6^j$ are the epochs' affine transformation parameters for the y-axis.
    \item $c_j$ is the epoch Differential Chromatic Refraction (DCR).
\end{itemize}

For convenience, we group the fitted parameters into two vectors:
\begin{enumerate}
    \item Source-related parameters: $\mathbf{s} = \left(x_0, y_0, \mu_x,\mu_y, \pi \right)$, where bold faces indicates vectors.
    \item Per-epoch parameters: $\mathbf{e} = \left(\mathbf{a}, c\right)$, where $\mathbf{a}$ collects the affine coefficients.
\end{enumerate}

\subsection{proper motion and parallax}
\label{sec:source_model}
The apparent motion of a stellar object in the sky can be described by
\begin{align}
    x_i(t_j)= x_i^0 + \mu_{x,i} t_j +\pi_i \bar{\omega}_x(t_j)\\
    y_i(t_j)= y_i^0 + \mu_{y,i} t_j +\pi_i \bar{\omega}_y(t_j)
\end{align}
where $x_i^0$  and $y_i^0$ are the positions in a reference time ($t=0$) in the $x$ and $y$ directions, $\mu_{x,i}$ and $\mu_{y,i}$ are the proper motions, ${\bar{\omega}_x}(t)$ (${\bar{\omega}_y}(t)$) is the parallax apparent motion function on the sky, with unity amplitude, projected on the x (y)-axis, and $\pi_i$ is the scale factor (i.e., the parallax measure). The apparent parallax motion in $\alpha$ and $\delta$ (i.e., right ascension and declination, respectively) is written by
\begin{align}
    \bar{\omega}_\alpha(t_j) &= X(t_j)\sin(\alpha) - Y(t_j)\cos(\alpha),\\
    \bar{\omega}_\delta(t_j) &= X(t_j)\cos(\alpha)\sin(\delta) + Y(t_j)\sin(\alpha)\sin(\delta) - Z(t_j)\cos(\delta) ,\nonumber
\end{align}
where $X(t_j),Y(t_j),Z(t_j)$ are the Barycentric coordinates of Topocentric observer at time $t_j$. 
Notice that for the KMTNet orientation $x\parallel -\alpha$ and $y \parallel \delta$.

In practice, however, we do not fit parallax to the real data analysed in this work, as the annual effect introduces systematics that dominate over the expected parallax signal 
(see \S\ref{sec:annual_effects_plx}).

\subsection{Affine Transformation}
\label{sec:AffineTransformation}

The affine transformation applied at each epoch maps the reference-frame source coordinates \( (x^0, y^0) \) to observed coordinates \( (x', y') \), and can be written in homogeneous form as:
\begin{align}
    \begin{bmatrix}
        x'\\y'\\1
    \end{bmatrix}
    =
    \begin{bmatrix}
        a_{11} & a_{12} & a_{13} \\
        a_{21} & a_{22} & a_{23} \\
        0 & 0 & 1
    \end{bmatrix}
    \begin{bmatrix}
        x^0 \\ y^0 \\ 1
    \end{bmatrix},
    \label{eq:affine_model}
\end{align}
where the \( a_{kl} \) are the affine transformation parameters for the given epoch, and \( (x^0, y^0) \) are the source positions in the reference frame.

This expression is formally non-linear in the unknowns because both the transformation parameters and the source positions are typically fitted. However, in our iterative solution, we alternate between solving for source positions and affine parameters while holding the other fixed. Under this scheme, Equation~\ref{eq:affine_model} becomes linear in each parameter.

In this work, we restrict the transformation to an affine model, as we operate on small cutouts with a relatively small field of view in which the distortions are expected to be minimal. However, the method described here is general and can accommodate higher-order transformations if needed for larger or more distorted fields, especially for wider fields of view or higher zenith angle, where atmospheric differential refraction is significant. For our fields of view, this effect is $\mathord{\sim} 10^{-2}\,\mathrm{mas}$ (using the formula from \citealt{seidelmann1992explanatory}), so an affine transformation is sufficient.
However, this should be done with care, as this may introduce new degeneracies into the model. 

\subsection{Differential Chromatic Refraction}\label{sec:chromatic_effect_model}

The amplitude of Differential Chromatic Refraction (DCR), $c$, depends on the atmospheric conditions at each epoch, specifically the temperature ($T$), pressure ($P$), and humidity ($H$), and can be expressed as a function of the form $c(T, P, H)$.
For broad-band, seeing-limited observations, the chromatic shift introduced by DCR can be approximated as:
\begin{align}
    \Delta \vec{x} = c\,\bar{C}\,\sec{z}\,(\hat{\text{pa}} \cdot \hat{x}), \label{eq:chromatic_reff}
\end{align}
where $\vec{x}=(x,y)$, $C$ is the source color, $\bar{C} \equiv C - \mathrm{median}(C)$ is the color relative to the field median, $\sec{z}$ is the airmass, $\hat{\text{pa}}$ is the unit vector along the parallactic angle, and $\hat{x}$ is the direction of the astrometric shift. The parameter $c$ represents the amplitude of the chromatic effect.
In the case of KMTNet, we expect the induced positional shift to be roughly proportional to $\Delta \delta \mathord{\propto} \cos(\text{pa})$ and $\Delta \alpha\, \mathord{\propto}\mathord{-} \sin(\text{pa})$.

Because the DCR-induced shift is typically small (on the order of $\lesssim 10$\,mas), it is not practical to fit this effect separately for every source and epoch. Instead, we group sources into color bins and model the DCR within each bin, using a weighted sum to calculate the effective residuals and weight, as described below. Specifically, sources are binned according to their $V-I$ color, using a bin width of 0.5\,mag, resulting in six color bins. 
The impact and correction of this effect are discussed in \S\ref{sec:chromatic_correction} and illustrated in Figures~\ref{fig:chromatic_X_axis_17K0103} and~\ref{fig:chromatic_Y_axis_17K0103}. In \S\ref{sec:matrix_form}, we explain in detail the matrix form that we use to fit the DCR effect.

\section{Minimisation}
\label{sec:minimization}
We follow the Astrometric Global Iterative Solution (AGIS, \citealt{lindegren2012astrometric}), with some modifications, for the astrometry solution and minimisation. 
The procedure is mathematically equivalent to the Gauss–Newton method for non-linear least-squares, 
implemented in block form to separately update source, affine, and DCR parameters.
Because we have modified the algorithm to include the colour terms and additional detrending steps, we write the algorithm explicitly for our case here.

The minimisation problem can be written as 
\begin{align}
    \min_{\mathbf{s},\mathbf{e}} \,\,Q & = \sum_{ {i,j}} R_{ij}^2 w_{ij},
    \label{eq:min_Q}
    \end{align}
    where $w_{ij}$ is a weight factor of the $ij$ observation (see \S\ref{sec:weights}), $R_{ij}$ is the residuals between the observations and the model, written as:
    \begin{align}
    R_{ij} &= \tilde{x}_{ij} - x_{ij}(\mathbf{s},\mathbf{e}),
\end{align}
  where $\tilde{x}_{ij}$ is the observed position of source $i$ at epoch $j$, and $x_{ij}(s,e)$ is the $i$ source model at epoch $j$.

Throughout the text, we denote scalar weights with lowercase $w$, and use uppercase $W$ to refer to the corresponding diagonal weight matrices in the linear algebra formulation (See~\S\ref{sec:matrix_form}).

To minimise Equation \ref{eq:min_Q}, we need to solve 
\begin{align}
    0=\frac{\partial Q}{\partial \varphi} &= 2\sum_{i,j} R_{ij}\frac{\partial{R_{ij}}}{\partial \varphi}w_{ij},
\end{align}
where $\varphi$ is a dummy parameter that indicates the fitted unknown parameters.
In our case, we assume we have a sufficient initial guess to neglect higher terms in the Taylor expansion. 
Therefore, the Taylor expansion around $\varphi^0$ can be written as
\begin{align}
    R_{ij}(\varphi^0+\epsilon) &\approx R_{ij}(\varphi^0) +\frac{\partial{R_{ij}}}{\partial\varphi} (\varphi^0) \epsilon + \mathord{O}(\epsilon^2).
    \label{eq:taylor_exp}
\end{align}
Here $\epsilon$ is a small perturbation vector relative to $\varphi^0$, representing the deviation from the initial guess.
Inserting Equation \ref{eq:taylor_exp} in Equation \ref{eq:min_Q}:
\begin{align}
   Q   = \sum_{ij} \left(R_{ij}(\varphi^0) + \frac{\partial{R_{ij}}}{\partial\varphi}  \epsilon \right) \left(R_{ij}(\varphi^0) + \frac{\partial{R_{ij}}}{\partial\varphi'}  \epsilon' \right) w_{ij},
\end{align}
\begin{align*}
    Q &= \sum_{ij} R_{ij}(\varphi^0)^2w_{ij}  + \frac{\partial{R_{ij}}}{\partial\varphi}  \epsilon \frac{\partial{R_{ij}}}{\partial\varphi'}  \epsilon' w_{ij}  \\ 
   & \quad + R_{ij}(\varphi^0)\left(\frac{\partial R_{ij}}{\partial \varphi} R_{ij}(\varphi^0)  \epsilon + \frac{\partial R_{ij}}{\partial \varphi'} R_{ij}(\varphi^0)  \epsilon' \right) w_{ij}.\nonumber 
\end{align*}
Here $\epsilon'$ is a small perturbation vector relative to $\varphi'^0$.
Since $Q$ is a scalar quadratic function of $\epsilon$ and $\epsilon'$, and the cross-term, $\left(\partial R_{ij}/\partial \varphi\right) \epsilon \cdot \left(\partial R_{ij}/\partial \varphi'\right)\epsilon'$, is symmetric under exchange of $(\varphi, \epsilon) \leftrightarrow (\varphi', \epsilon')$, we may, without loss of generality, set $\epsilon' = \epsilon$ in order to compute the gradient with respect to $\epsilon$. Minimizing $Q$ with respect to $\epsilon$ (i.e., setting $\partial Q / \partial \epsilon = 0$) then leads to the normal equations:
\begin{align}
    \sum_{ij}  \frac{\partial{R_{ij}}}{\partial\varphi}   \frac{\partial{R_{ij}}}{\partial\varphi'}w_{ij} \epsilon  &=  - R_{ij}(\varphi^0) \frac{\partial R_{ij}}{\partial \varphi}  w_{ij}.
    \label{eq:normalequation}
\end{align}
This can be written in a matrix form like:
\begin{align}
    Nu=b
    \label{eq:normalequation_matrix}
\end{align}
The matrix $N$, and vectors $u$ and $b$, are constructed from sub-components:
\begin{align}
    \begin{bmatrix}
        N_{ss} & N_{se} \\
        N_{es} & N_{ee} \\
    \end{bmatrix} 
    \begin{bmatrix}
         u_{s}\\
         u_{e}\\
    \end{bmatrix}
    = 
    \begin{bmatrix}
         b_{s}\\
         b_{e}
    \end{bmatrix},
    \label{eq:matrix_eq}
\end{align}
where the subscripts $s$ and $e$ denote derivatives with respect to the source and epoch parameters, respectively. 
For clarity, we further decompose the per-epoch vector as $\mathbf{e} = (\mathbf{a},\, c)$ when discussing the affine and DCR sub-components in \S\ref{sec:matrix_form}.
One important result is that in the case of a linear model, only $b$ is a function of the free parameters and the residuals.
Therefore, we only need to update $b_s$ and $b_e$, keep $u$, and recalculate the residuals in each iteration. The linear systems were solved using the bi-conjugate gradients method\footnote{Using the built-in {\tt bicg} function in MATLAB.}. We define a step as a full update of the model parameters (source, epoch, chromatic terms), and an iteration as the internal loop solving the normal equations (Equation \ref{eq:normalequation}).

\subsection{Uncertainties and weights}
\label{sec:weights}

We assign weights (via the matrix $W$) to individual observations per epoch and source based on empirical residuals. Each star is assigned a weight derived from the astrometric scatter of stars with similar magnitude and the same epoch.

Because the uncertainties are estimated from the residuals and the solution is obtained iteratively, we assume uniform weights for all observations in the first step. In the following steps, we compute a magnitude-dependent astrometric scatter by applying a median filter (with a window size of 1 magnitude) to the 2D astrometric residuals. Based on this, we assign a weight to each source of magnitude $m$ at epoch $j$ according to:
\begin{equation}
    w_{m,j} = \frac{1}{\sigma^2_{i\in m,j}},
\end{equation}
where $\sigma_{i\in m,j}$ is the median 2D astrometric scatter for stars belonging to magnitude bin $m$, computed as described above.

To identify and suppress outliers, we use a similar approach. Specifically, we apply a running median and a median absolute deviation (MAD) filter to the 2D astrometric scatter as a function of magnitude. Sources whose RMS exceeds three times the local MAD plus the local median are flagged as outliers, and their weights are reduced by a factor of ten. We adopt this mild down-weighting as a {\it heuristic compromise}: it suppresses their influence while keeping them in the fit, because in an iterative scheme, some measurements can switch between inlier and outlier status across iterations.

\begin{figure*}
  \centering
  \begin{subfigure}[t]{0.48\textwidth}
    \includegraphics[width=\textwidth]{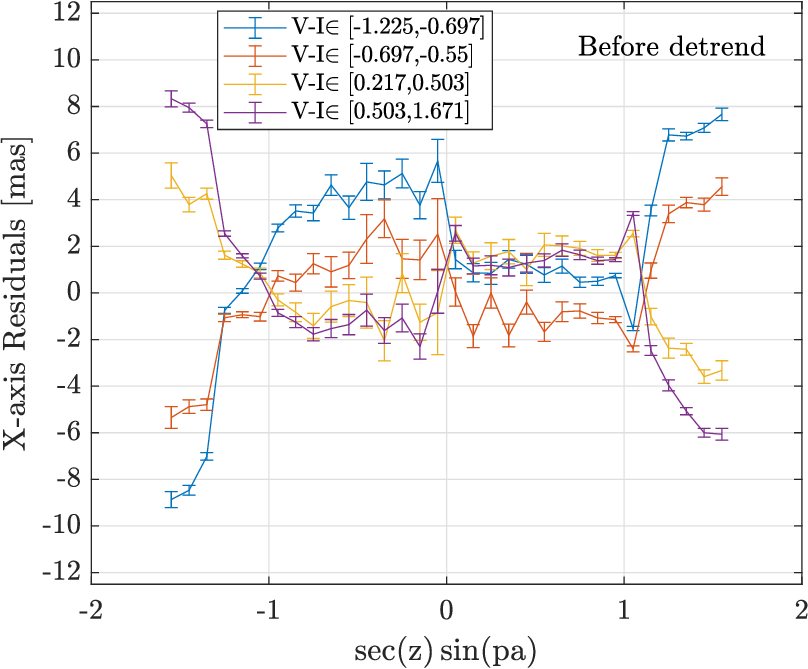}
    \caption{}
    \label{fig:chromatic:a}
  \end{subfigure}
  \hfill
  \begin{subfigure}[t]{0.48\textwidth}
    \includegraphics[width=\textwidth]{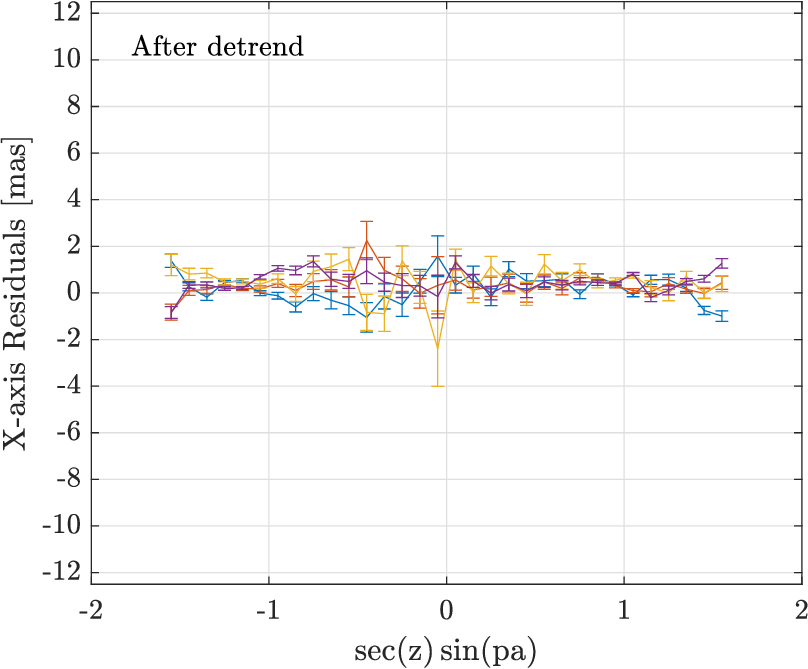}
    \caption{}
    \label{fig:chromatic:b}
  \end{subfigure}
  
  \begin{subfigure}[t]{0.48\textwidth}
    \includegraphics[width=\textwidth]{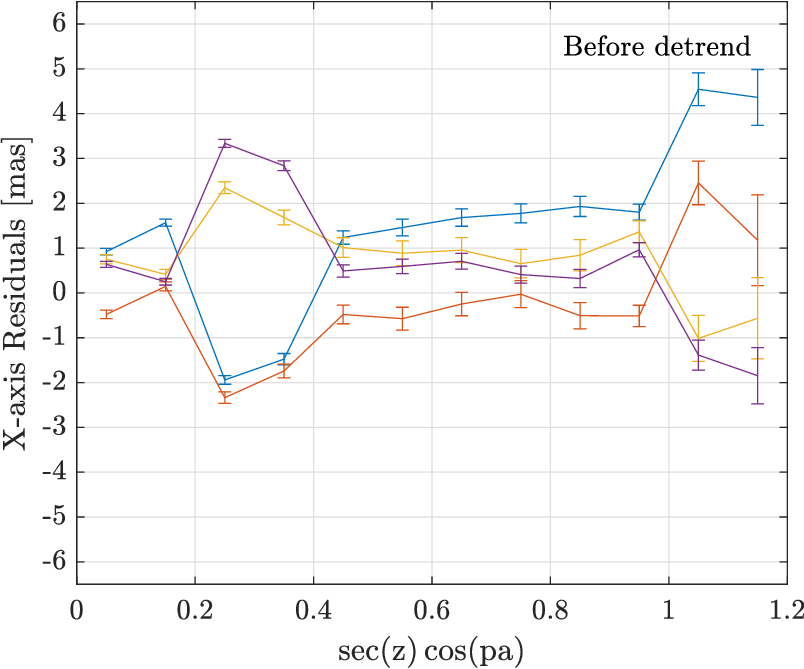}
    \caption{}
    \label{fig:chromatic:c}
  \end{subfigure}
  \hfill
  \begin{subfigure}[t]{0.48\textwidth}
    \includegraphics[width=\textwidth]{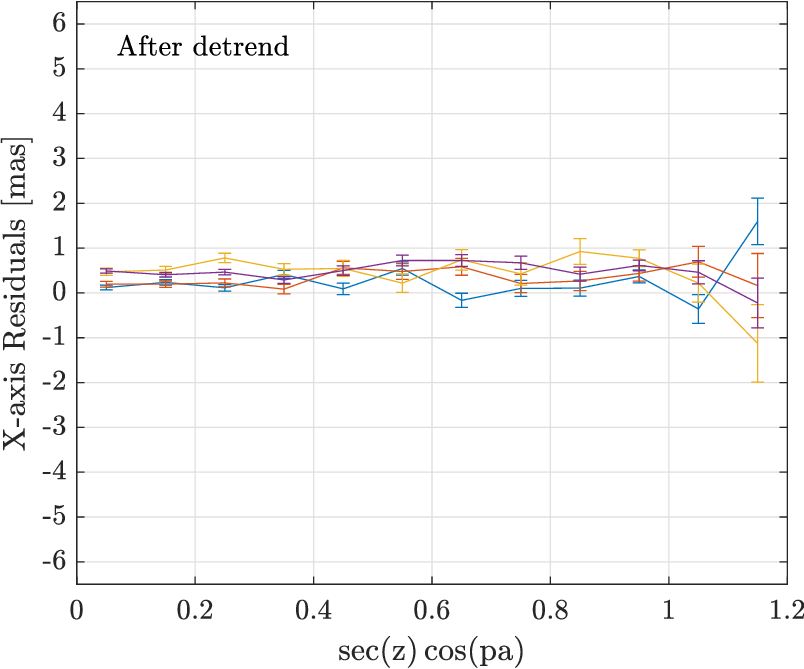}
    \caption{}
    \label{fig:chromatic:d}
  \end{subfigure}

  \caption{Astrometric residuals as a function of the DCR-leading terms in the x-axis. Left panels show the residuals without DCR correction, while right panels show the residuals after applying the correction. Error bars represent the mean and standard error of the mean (i.e., standard deviation divided by the square root of the number of measurements) in each bin of $V-I$ color and corresponding DCR term. Only the top 33\% of sources, based on 2D single-epoch precision ($\leq 10$\,mas), are included.}
  \label{fig:chromatic_X_axis_17K0103}
\end{figure*}

\begin{figure*}
  \centering
  \begin{subfigure}{0.48\textwidth}
    \includegraphics[width=\textwidth]{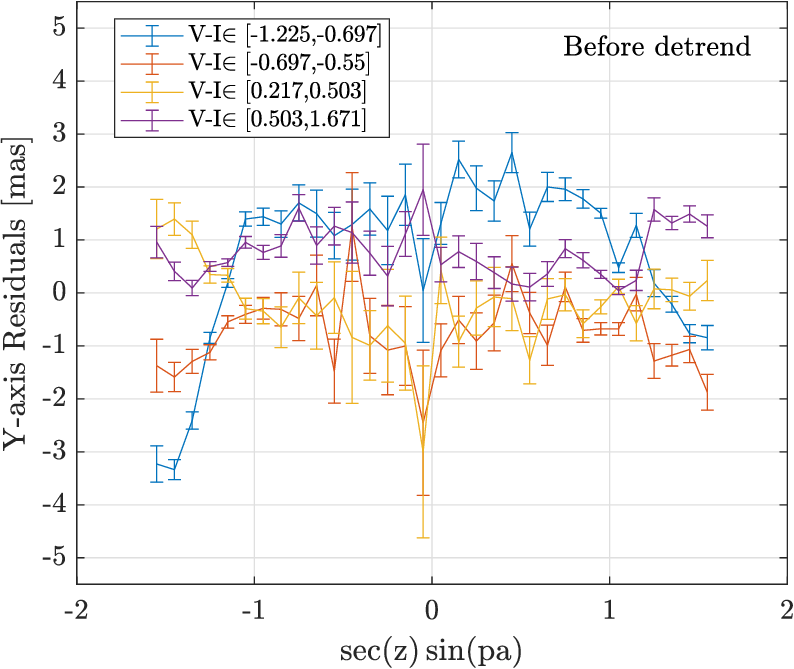}
    \caption{}
    \label{fig:chromaticY:a}
  \end{subfigure}
  \hfill
  \begin{subfigure}{0.48\textwidth}
    \includegraphics[width=\textwidth]{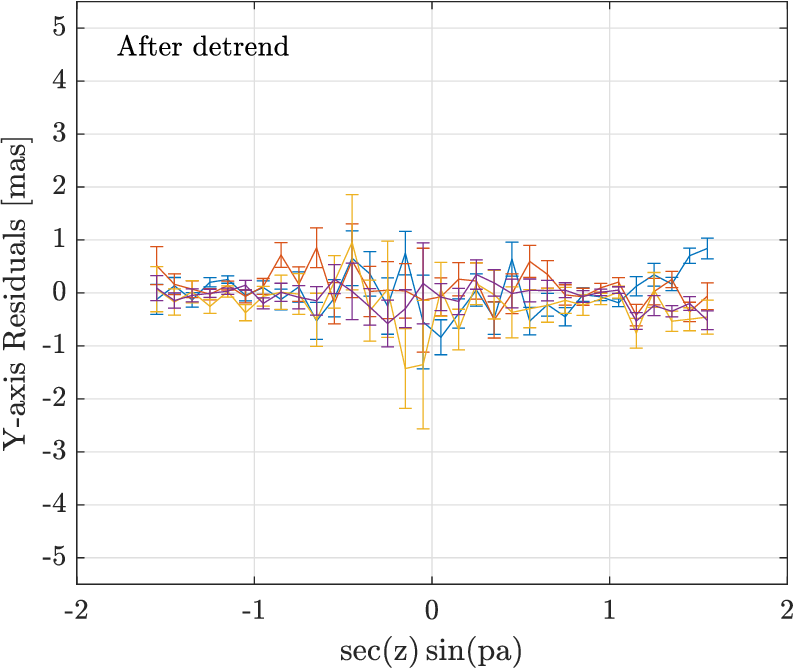}
    \caption{}
    \label{fig:chromaticY:b}
  \end{subfigure}
  \vspace{1ex}

  \begin{subfigure}{0.48\textwidth}
    \includegraphics[width=\textwidth]{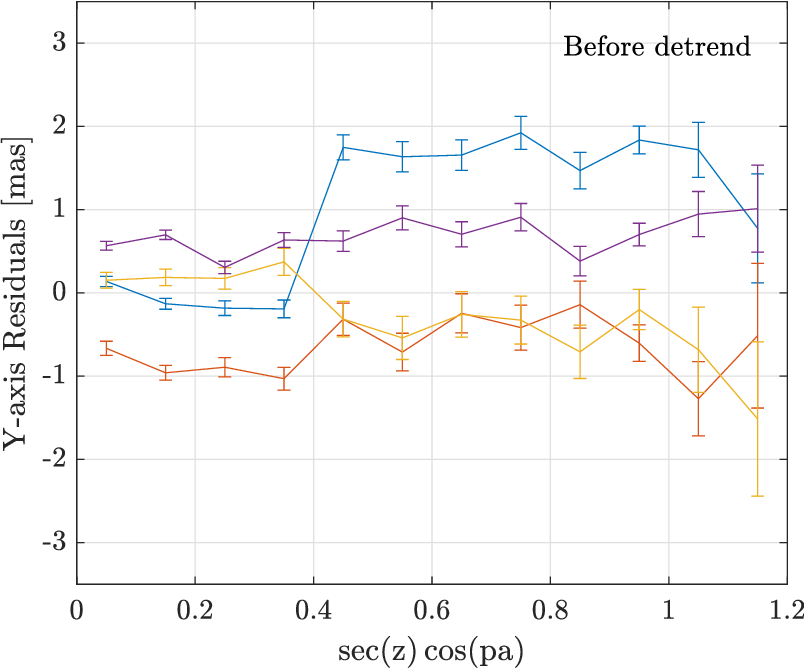}
    \caption{}
    \label{fig:chromaticY:c}
  \end{subfigure}
  \hfill
  \begin{subfigure}{0.48\textwidth}
    \includegraphics[width=\textwidth]{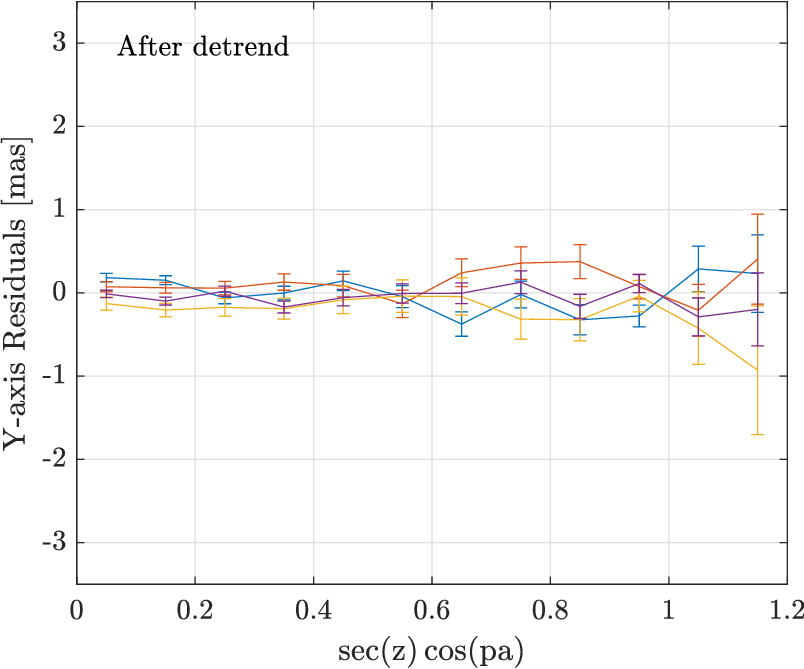}
    \caption{}
    \label{fig:chromaticY:d}
  \end{subfigure}

  \caption{Same as Figure~\ref{fig:chromatic_X_axis_17K0103}, but for the y-axis.}
  \label{fig:chromatic_Y_axis_17K0103}
\end{figure*}

\section{linear algebra formulation}
\label{sec:matrix_form}

In this section, we present the matrix formulation of the full astrometric model, which includes source parameters (initial position and proper motion), epoch-dependent affine transformations, and Differential Chromatic Refraction (DCR). The model is solved using the iterative weighted least-squares approach described in \S\ref{sec:minimization}.

Each group of parameters (source proper motion, affine transformation, and DCR) is associated with its own design matrix. We begin by describing the design matrices and normal equations related to the source proper motion and affine transformation parameters, and then extend the formulation to include the DCR model.

\subsection{Source and Affine Parameters}

We consider a 2D astrometric model that includes affine transformations and source motion, excluding parallax. The model is structured such that each source and each epoch contributes independently to the design matrices, which we define below.

The source model accounts for reference position and linear motion (proper motion) over time. For a single source observed over multiple epochs, the corresponding design matrices are:
\begin{align}
    H_{s,x} &= [1, 0, \bar{t}_j, 0, \bar{\omega}_{x}(t_j)], \\
    H_{s,y} &= [0, 1, 0, \bar{t}_j,\bar{\omega}_{y}(t_j)],
    \label{eq:source_design}
\end{align}
where the columns of these matrices (left to right) correspond to the reference position in $x$ ($x_0$), reference position in $y$ ($y_0$), proper motion in $x$ ($\mu_x$), proper motion in $y$ ($\mu_y$), and the parallax ($\pi$). The relative time variable is defined as \( \bar{t}_j \equiv t_j - t_0 \), where \( t_0 \) is a fixed reference epoch\footnote{In practice $t_0$ is selected to be the mid observing time.}, and $\bar{\omega}_x(t_j)$ is the parallax apparent motion projection on the x axis.
The source design matrices \( H_s \) have dimensions \( N_{\mathrm{epoch}} \times 5 \). However, in this work, we exclude the parallax, as discussed in \S\ref{sec:annual_effects_plx}.

The affine model describes the per-epoch linear transformation that maps reference-frame coordinates to detector coordinates. For a single epoch, the design matrices are:
\begin{align}
    H_{a,x} &= [x_0, y_0, 1, 0, 0, 0], \\
    H_{a,y} &= [0, 0, 0, x_0, y_0, 1],
    \label{eq:epoch_design}
\end{align}
where \( x_0, y_0 \) are the source coordinates at the reference epoch.  
The affine design matrices \( H_e \) have dimensions \( N_{\mathrm{src}} \times 6 \), which correspond to the six parameters of the affine transformation on the two axes, as described in Equation \ref{eq:affine_model}.

The normal matrix and right-hand side vector for the source parameters are given by:
\begin{align}
    N_{ss} &= H^\intercal_{s,x} W_e H_{s,x} + H^\intercal_{s,y} W_e H_{s,y}, \\
    b_s &= H^\intercal_{s,x} W_e R_x + H^\intercal_{s,y} W_e R_y,
\end{align}
where \( R_x \) and \( R_y \) are the residual vectors for the \(x\) and \(y\) coordinates, of size \( N_{\mathrm{epoch}} \times N_{\mathrm{src}} \), and \( W_e \) is a diagonal matrix encoding the weights (inverse variances) for each epoch.

Similarly, the normal matrix and right-hand side for the epoch (affine) parameters are:
\begin{align}
    N_{aa} &= H^\intercal_{a,x} W_s H_{a,x} + H^\intercal_{a,y} W_s H_{a,y}, \\
    b_a &= H^\intercal_{a,x} W_s R_x + H^\intercal_{a,y} W_s R_y,
\end{align}
where \( W_s \) is a diagonal matrix of size \( N_{\mathrm{src}} \times N_{\mathrm{src}} \), representing inverse variance weights for the sources.

\subsubsection{Differential Chromatic Refraction (DCR).}

To account for chromatic shifts due to the Earth's atmosphere, we add the per-epoch chromatic term $c$ to the model, 
as described in \S\ref{sec:chromatic_effect_model}.

For each color bin, we define a common design matrix \( H_c \) of size \( N_{\mathrm{epoch}} \times 8 \), which captures the angular dependence of the chromatic effect up to the fourth order:

\begin{align}
    H_c &= \sec{z} \cdot \big[
    \sin(\mathrm{pa}), \cos(\mathrm{pa}),\sin^2(\mathrm{pa}), \cos^2(\mathrm{pa}),\\ \nonumber 
    &\sin^3(\mathrm{pa}), \cos^3(\mathrm{pa}),\sin^4(\mathrm{pa}), \cos^4(\mathrm{pa})\big].
\end{align}

The normal matrix and right-hand side for the DCR parameters in a given color bin $B$ are constructed as:
\begin{align}
    N_{cc} &= H_c^\intercal W_c H_c, \\
    b_c &= \sum_{j} H_j^\intercal \left( \sum_{i \in B} W_{i,j} R_{i,j} \right),
\end{align}
where \( R_{i,j} \) is the residual position of source \( i \) at epoch \( j \) (after subtracting other model components), and the diagonal weight matrix \( W_c \) is defined per epoch as:
\begin{align}
    W_{c,j} = \sum_{i,\,c_i \in B} W_{i,j}.
\end{align}
Here, \( c_i \) indicates the color bin membership of source \( i \), and \( W_{i,j} \) is the weight of source \( i \) at epoch \( j \).

Solving the normal equation,
\begin{equation}
    N_{cc} x_c = b_c,
\end{equation}
yields the DCR coefficients $x_c$, which are then used to compute the chromatic correction for all sources within the bin. 
This model neglects chromatic aberrations due to the telescope optics. The rationale is that, since in most cases, the stars are placed in roughly the same position in the focal plane, then, to the first order, such aberrations are roughly constant.

\begin{figure*}
  \centering
  \begin{subfigure}[t]{0.48\textwidth}
    \includegraphics[width=\textwidth]{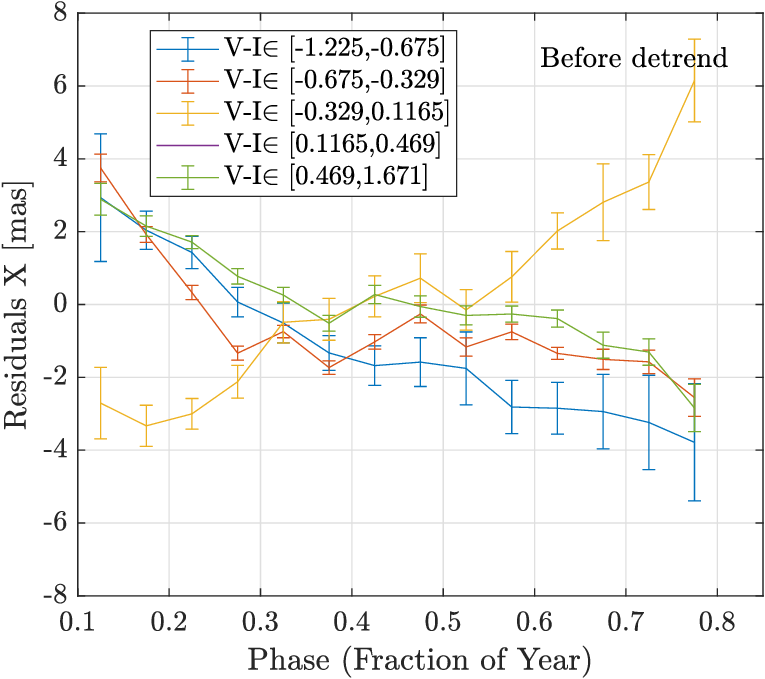}
    \caption{}
    \label{fig:annual:a}
  \end{subfigure}
  \hfill
  \begin{subfigure}[t]{0.48\textwidth}
    \includegraphics[width=\textwidth]{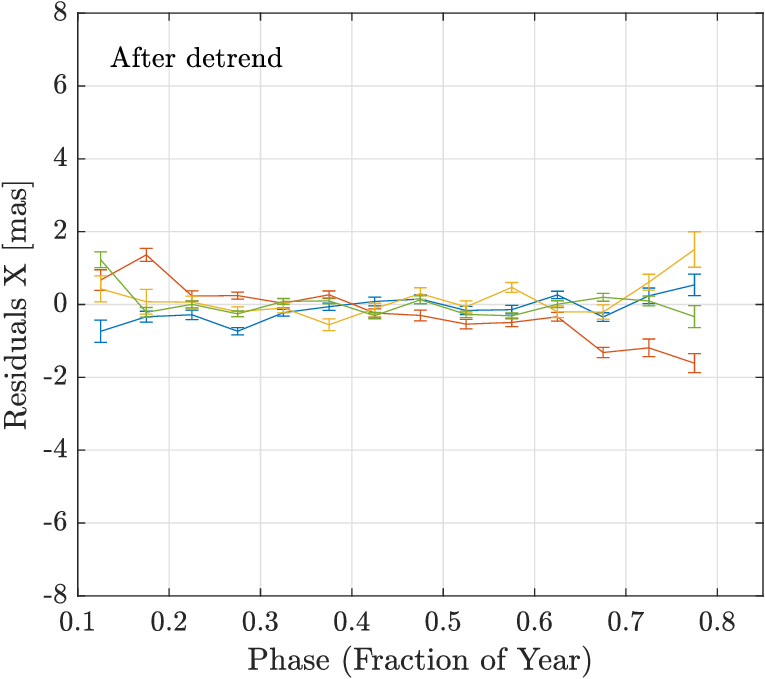}
    \caption{}
    \label{fig:annual:b}
  \end{subfigure}
  \vspace{1ex}

  \begin{subfigure}[t]{0.48\textwidth}
    \includegraphics[width=\textwidth]{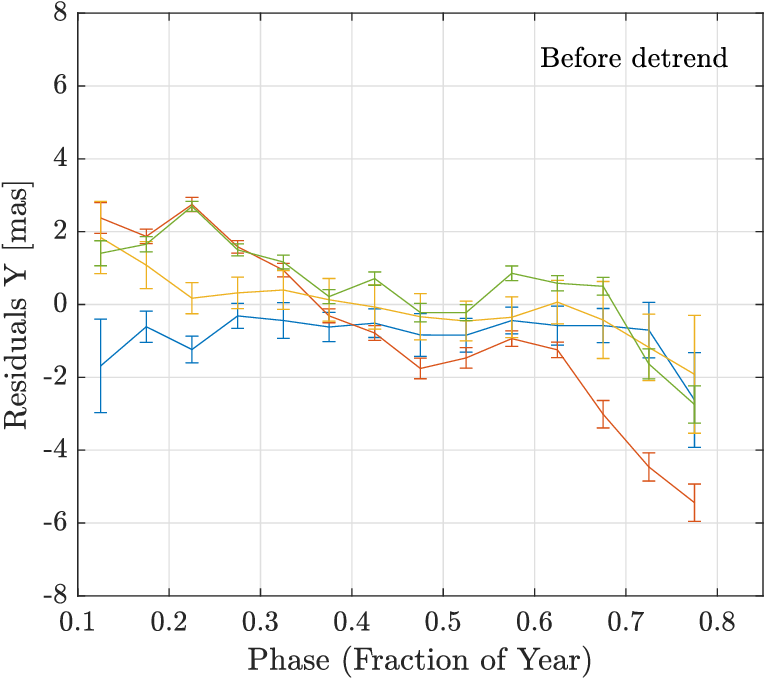}
    \caption{}
    \label{fig:annual:c}
  \end{subfigure}
  \hfill
  \begin{subfigure}[t]{0.48\textwidth}
    \includegraphics[width=\textwidth]{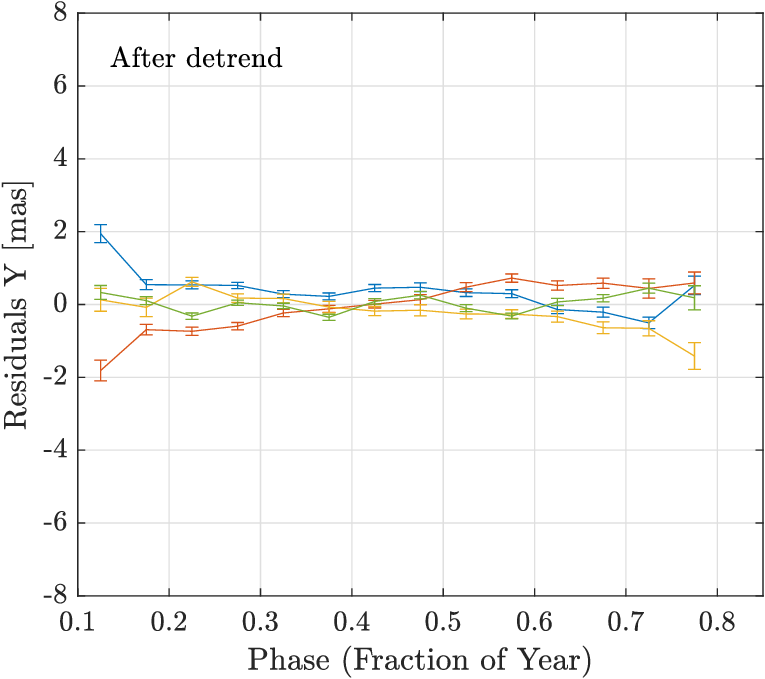}
    \caption{}
    \label{fig:annual:d}
  \end{subfigure}

  \caption{Astrometric residuals as a function of the annual phase for both axes, before (left) and after (right) detrending. Residuals are binned in intervals of 0.05 in phase (fraction of year), and colored by $V-I$ color bins as indicated in the legend. Each point represents the median residual within a bin, and the error bars show the standard error. We use the top 33\% best sources, with 2D single-epoch precision of \(\leq 10\,\mathrm{mas}\).}
  \label{fig:annual_effect}
\end{figure*}

\begin{figure}\centering
\includegraphics[width=1\linewidth]{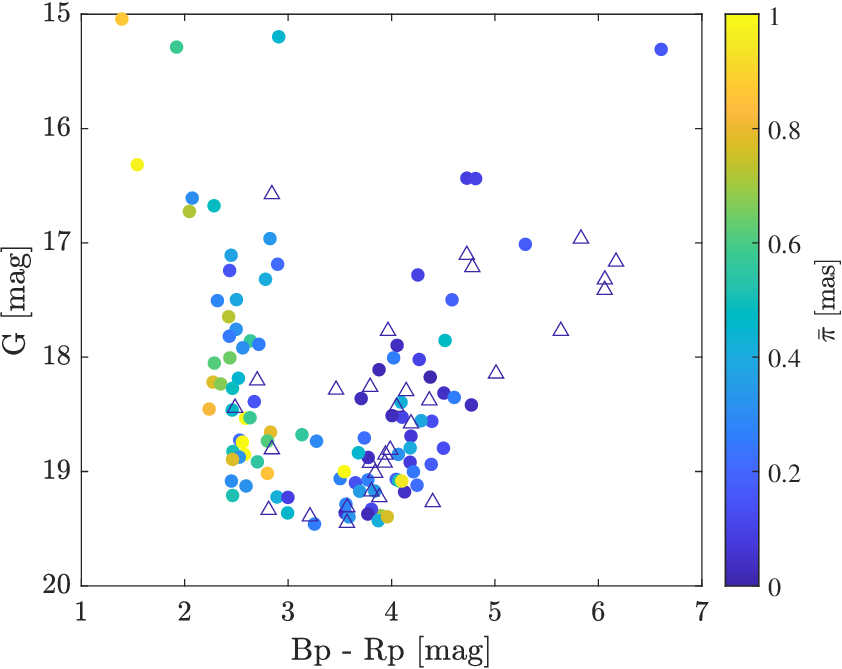}
\caption{Colour–magnitude diagram (CMD) based on Gaia DR3 parameters for a 300$\times$300 pixel cutout of the KMTNet BLG17K0103 field. Colours indicate Gaia DR3 parallax values, with hollow triangles denoting sources with negative parallax.}
\label{fig:cmd_192630}
\end{figure}

\subsection{Simple Example}

To illustrate the matrix formulation described in \S\ref{sec:matrix_form}, we present a simplified example with two sources and two observations. This minimal case is intended to clarify the structure and coupling of the source and epoch matrices in the least-squares system. For simplicity, we restrict the model to a single axis (x), exclude parallax and DCR, and assume linear motion plus a first-order affine transformation per epoch. We also assume all observations are equally weighted, i.e., the weight matrix is the identity matrix and all residuals contribute equally to the fit.

In this simplified example, the per-epoch parameters reduce to the affine coefficients only, so we denote them by $\mathbf{a}$ and write the normal equations using $a$–subscripts.

We adopt the following astrometric model:
\begin{align}
    x_{ij} = x_i^0 + \mu_i t_j + a^j_1 x_i^0 + a^j_2 y_i^0 + a^j_3,
    \label{eq:affine_source}
\end{align}
where \( i \in \{1,2\} \) indexes sources and \( j \in \{1,2\} \) indexes epochs.

The source design matrix \( H_{s,x} \) includes one row per epoch and has size \( N_{\mathrm{epoch}} \times 4 \). Each row corresponds to:
\begin{align}
    H_{s,x}^{(j)} = [1,\, 0,\, \bar{t}_j,\, 0],
\end{align}
where the four components correspond to the reference position in x, the reference position in y, the proper motion in x, and the proper motion in y. The time variable \( \bar{t}_j \equiv t_j - t_0 \) denotes the time relative to a fixed reference epoch \( t_0 \).
The full normal matrix for the source parameters is computed as:
\begin{align}
    N_{ss} = H_{s,x}^\intercal H_{s,x} = \sum_j H_{s,x}^{(j)\,\intercal} H_{s,x}^{(j)}.
\end{align}

To illustrate the structure, we now break down the contribution from a single epoch \( j \):
\begin{align}
    N_{ss}^{(j)} =
    H_{s,x}^{(j)\,\intercal} H_{s,x}^{(j)} =
    \begin{bmatrix}
        1 & 0 & \bar{t}_j & 0 \\
        0 & 0 & 0 & 0 \\
        \bar{t}_j & 0 & \bar{t}_j^2 & 0 \\
        0 & 0 & 0 & 0 \\
    \end{bmatrix}.
\end{align}
Since this example models only the x-coordinate, the components related to y-position and y-proper-motion (second and fourth rows/columns) are structurally zero.

The coupling between source and epoch parameters is captured in the off-diagonal block \( N_{se} \) (and its transpose \( N_{es} = N_{se}^\intercal \)):
\begin{align}
    N_{es} = N_{se}^\intercal =
    \begin{bmatrix}
        x_i^0 a^1_j & x_i^0 a^2_j & 0 & 0 \\
        y_i^0 a^1_j & y_i^0 a^2_j & 0 & 0 \\
        a^1_j       & a^2_j       & 0 & 0 \\
    \end{bmatrix}.
\end{align}
The epoch block \( N_{ee} \) takes the form:
\begin{align}
    N_{ee} =
    \begin{bmatrix}
        (x_i^0)^2     & x_i^0 y_i^0     & x_i^0 \\
        x_i^0 y_i^0   & (y_i^0)^2       & y_i^0 \\
        x_i^0         & y_i^0           & 1
    \end{bmatrix}.
\end{align}
The complete normal matrix system for this case has the following block structure, where the first two rows and columns correspond to the sources, and the last two to the epochs:
\begin{align}
    \begin{bmatrix}
        N_{ss}^{11} + N_{ss}^{12} & \emptyset & N_{se}^{11^\intercal} & N_{se}^{12^\intercal} \\
        \emptyset & N_{ss}^{22} + N_{ss}^{21} & N_{se}^{21^\intercal} & N_{se}^{22^\intercal} \\
        N_{se}^{11} & N_{se}^{21} & N_{ee}^{11} + N_{ee}^{21} & \emptyset \\
        N_{se}^{12} & N_{se}^{22} & \emptyset & N_{ee}^{12} + N_{ee}^{22} \\
    \end{bmatrix}.
\end{align}

As shown by \cite{lindegren2012astrometric}, it is sufficient to neglect the off-diagonal matrices (e.g., \( N_{se} \)) in each iteration, as long as the parameters are updated sequentially. Based on this argument, we drop the cross-coupling terms and update the source and epoch parameters independently in each step. This allows us to solve the normal equations block by block: first solving for the source parameters using the current epoch solution, and then updating the epoch parameters using the updated source solution.

\section{Detrending}
\label{sec:detrending}

Real data may contain systematic effects that are not included in our initial model. We need a way to estimate these effects and detrend them.
To address this, we employ several basic detrending techniques. Specifically, we identify correlations between the astrometric residuals and various parameters, as well as applying the SYSREM algorithm (\citealt{sysrem_tamuz2005correcting}), which operates without requiring a model.
Table~\ref{tab:experiment_results} outlines the workflow for each algorithm and details the application of the detrending procedures.
In the following, we present the detrending methods we are using in the order we apply them.

\subsection{Chromatic effects}
\label{sec:chromatic_correction}
The refraction in the Earth's atmosphere (and sometimes in the telescope optics) is color-dependent, which may induce color-dependent astrometric variations.
Our basic chromatic model is included in the basic astrometric model and described in \S\ref{sec:chromatic_effect_model}.

To visualize the impact of this effect, we run our pipeline without incorporating the parallax and DCR models. 
The left panels of Figures \ref{fig:chromatic_X_axis_17K0103} and \ref{fig:chromatic_Y_axis_17K0103} illustrate the residuals along the X and Y axes, respectively, binned by color ($V-I$), as a function of the linear terms in the chromatic model (i.e., $AM\sin{q}$ and $AM\cos{q}$), where only sources with a 2D single epoch precision (RMS) of less than 14\,mas are presented.
Next, the right panels of Figures \ref{fig:chromatic_X_axis_17K0103} and \ref{fig:chromatic_Y_axis_17K0103} show the same after applying the chromatic model.
These plots demonstrate that for some stars the chromatic effects may be as large as $\sim10$\,mas, and that the inclusion of this effect in our astrometric model reduces the amplitude of the residuals to below $\approx1$\,mas.

\subsection{Annual effects}
\label{sec:annual_effects_plx}

Even after correcting for chromatic effects, the astrometric residuals exhibit systematic trends as a function of the day of the year. The left panels of Figure \ref{fig:annual_effect} show an example of this annual pattern, with residuals plotted against fractional year and binned by stellar color.
These trends are about an order of magnitude larger than the expected parallax of the stars in the field (Figure~\ref{fig:cmd_192630}), and are likely caused by seasonal variations in the Earth's atmosphere or local, site- or telescope-dependent systematics, that the basic chromatic model does not capture.
To correct this effect, we fit a 4th-order polynomial in the fractional year to the residuals within each color bin. This allows us to model smooth, color-dependent seasonal trends without fitting individual stars, which would likely risk removing real parallax signals. The fitted polynomial for each color bin is then subtracted from the corresponding sources' residuals.
As a result, the periodic trend is effectively removed from each color bin, leading to significantly reduced systematics, as shown in the right panels of Figure~\ref{fig:annual_effect}.
Although the fitted signal has a one-year period, its amplitude is typically $\sim$6\,mas—far larger than the expected parallax of sources in these fields (typically sub-mas). Moreover, the phase of this annual effect differs from that expected for parallax motion, and the color-bin-based correction confirms that it is more consistent with atmospheric refraction variations or local (site or instrumental) systematics than with astrophysical parallax.
However, due to the possible correlations between color and parallax (Fig.~\ref{fig:cmd_192630}) this approach may still remove the real parallax signal. Given this issue and the fact that the expected parallax signal in this field is of the order of our measurement errors, we do not attempt to fit the parallax signal.

\subsection{Intra-Pixel variation}
\label{subsec:intra_pixel}
Intra-pixel variation refers to systematic astrometric shifts that depend on a source's precise location within a detector pixel. These shifts may arise from various reasons
including:
sub-pixel sensitivity variations;
optical or electronic distortions across the pixel surface \citep{irac_astrometry_esplin2015measuring};
or Nyqusit under-sampled PSF (e.g., \citealt{Ofek2019_Astrometry_Code}).

To account for these effects, we model the intra-pixel shifts using a two-dimensional fifth-order polynomial in the sub-pixel coordinates. This polynomial is fitted to the residuals after all other model components have been subtracted, capturing systematic trends as a function of the source's position within each pixel. The resulting model is then subtracted from the residuals to remove intra-pixel systematics and improve the overall astrometric solution.

We did not detect a significant amplitude of this effect in our tests on the KMTNet images.

\subsection{SysRem}

SysRem \citep{sysrem_tamuz2005correcting} is a blind iterative detrending algorithm designed to identify and remove systematic effects, originally developed for photometric data.
This method works by approximating the residual matrix using the multiplication of two vectors (e.g., using Singular Value Decomposition, or iterative methods).
SysRem allows for removing systematic effects whose nature is unknown (a-priori).
For example, in photometric data, typically one SysRem vector will act as the zero point per image, while the second vector is
a zero point per star (e.g., due to color effects).
We never employ SysRem more than once. Otherwise, one can remove real variability from the data.

\begin{figure}
\centering
 \includegraphics[width=0.9\linewidth]{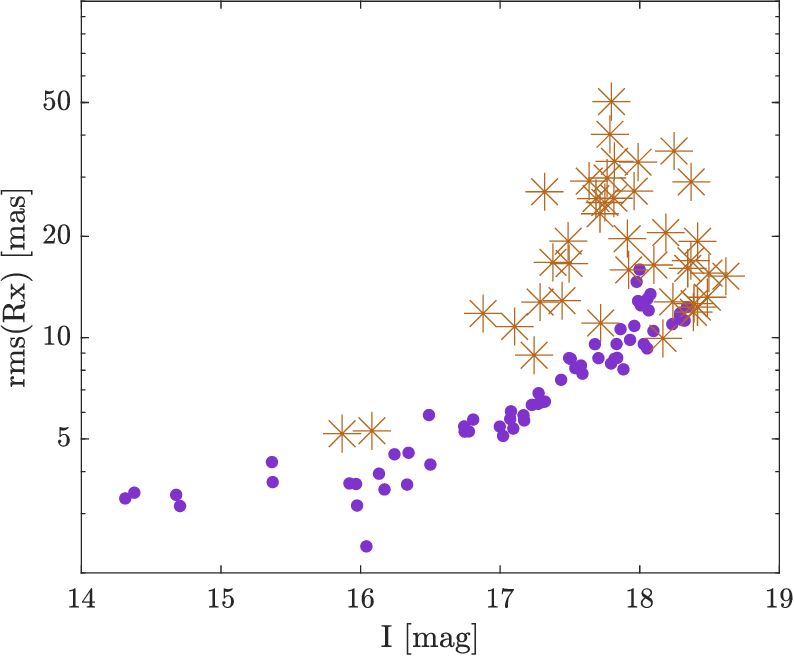}
 \includegraphics[width=0.9\linewidth]{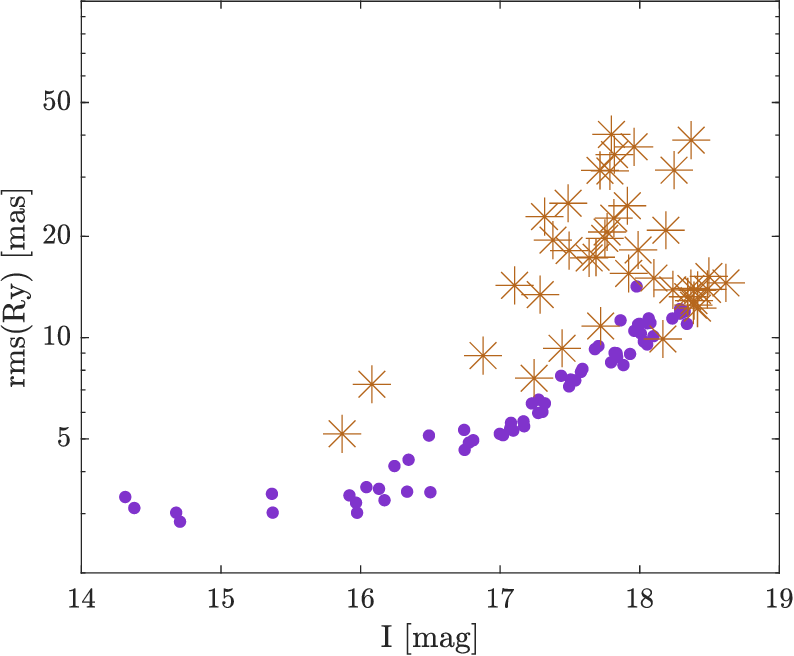}
 \includegraphics[width=0.9\linewidth]{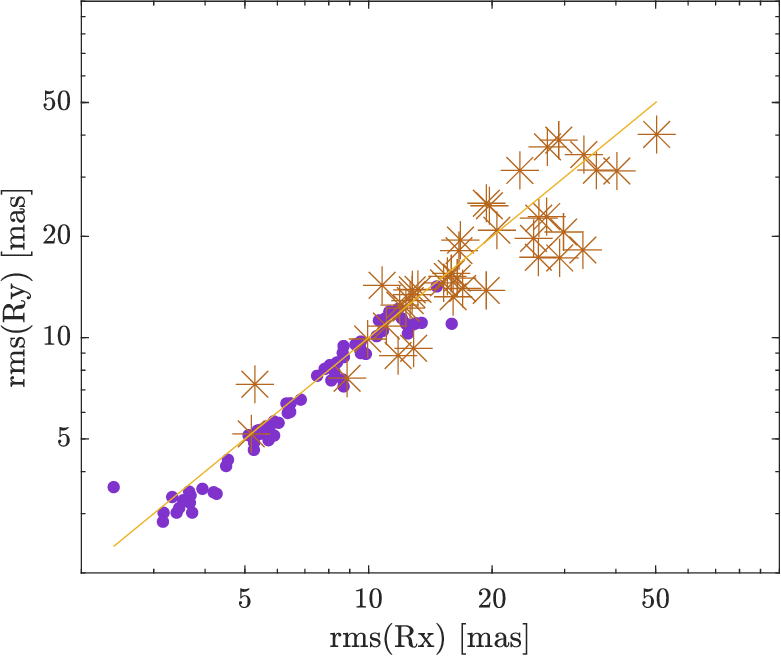}
 \caption{The astrometric single-epoch precision as a function of the source median magnitude. The precision is calculated as the root mean squared (rms) of the residuals around the best-fit model. Top: rms of the x-axis residuals (1D). Middle: rms of the y-axis residuals (1D). Bottom: rms of the y-axis residuals plotted against rms of the x-axis residuals, with $y = x$ line included for reference. The data points marked by brown asterisks are treated as outliers.}\label{fig:rms_vs_I_w}
 \end{figure}

\section{Step-by-Step Algorithm}
\label{sec:step_by_step_algo}

Here, we provide a detailed, step-by-step description of the full astrometric pipeline, including optional variants. Several steps, especially the detrending components, are heuristic in nature and may be applied at different stages. We have tested the pipeline under multiple configurations, which are summarised in Table~\ref{tab:experiment_results}. The version described below corresponds to algorithm 6 in Table~\ref{tab:experiment_results}, although the differences in precision among the various configurations are relatively small.

As outlined in \S\ref{sec:minimization}, the pipeline solves a linear model iteratively: we start with an initial solution and then refine the model parameters step by step through repeated cycles. This iterative structure allows us to decouple the various components of the astrometric model. In each iteration, we solve the Normal Equation (see Equation~\ref{eq:normalequation}) for a specific model component (e.g., proper motion, affine transformation, etc.), update the corresponding parameters, and then proceed to the next component in a fixed order. This modular approach follows the principles of the Astrometric Global Iterative Solution (AGIS) method developed for Gaia astrometry \citep{lindegren2012astrometric}, and facilitates improved convergence as well as reduced parameter correlations. We refer to such an iterative solution of Equation~\ref{eq:normalequation} as a "step". The first step is always performed without weights. Following this initial step, we perform additional steps to detrend the data and perform another step with weights.

The pipeline proceeds as follows:

\begin{itemize}
    \item \textbf{Match field catalogues:} For each observation epoch, we construct a combined catalogue containing the photometric and astrometric measurements of all sources. This results in a data matrix tracking all sources across epochs.
    
    \item \textbf{Align catalogues:} We fit an affine transformation between each epoch's catalogue and a common reference frame (based on the KMTNet catalogue). This transformation is applied to align the catalogues spatially, but we retain the original positions for later modelling.
    
    \item \textbf{Initial step:} We solve the system of equations (Normal Equation; Equation~\ref{eq:normalequation}) using equal weights for all observations. In this stage, we model the source proper motions and affine transformations for each epoch (see \S\ref{sec:source_model} and \S\ref{sec:AffineTransformation}). 
    Because the problem we are solving is not strictly linear, we apply multiple 'internal iterations' ($n^{int}_{iter}$), where in each iteration, we start with the predicted positions of the previous iteration.
    We refer to this as an “initial step” because it produces a first approximation of the model parameters before applying weights or detrending.
    
    \item \textbf{Weighted modelling step:} After the initial step, we refine the model with additional internal iterations. In each iteration, we compute updated residuals, assign new observation weights based on the empirical scatter (\S\ref{sec:weights}), fit the astrometric models, and apply the detrend procedures. This step also optionally includes the DCR component. Following each iteration, we optionally detrend for annual effects and intra-pixel variations.
    
    \item \textbf{SYSREM detrending:} After the modelling steps, we optionally run one iteration of the SYSREM algorithm \citep{sysrem_tamuz2005correcting} on the residuals. SYSREM models and removes common-mode systematics across sources and epochs.
    
    \item \textbf{Final refinement step:} We perform a final step that includes the source proper motions and affine transformation models and, optionally, the DCR. This ensures that any improvements from SYSREM are incorporated into the final astrometric solution.
\end{itemize}

This structured, iterative approach allows us to build a progressively more accurate astrometric model while systematically correcting for time-dependent and spatial systematics in the data.

\begin{figure}
\centering
 \includegraphics[width=0.9\linewidth]{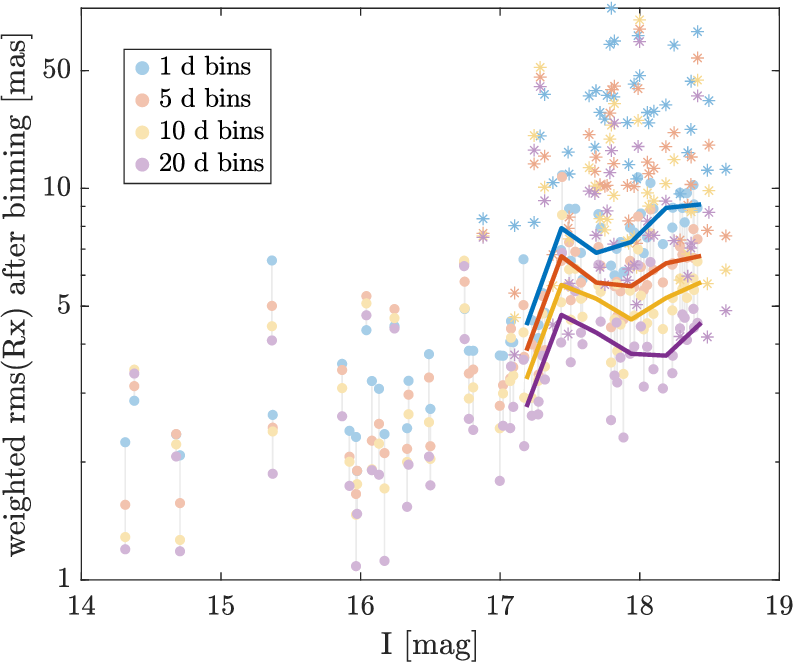}
 \includegraphics[width=0.9\linewidth]{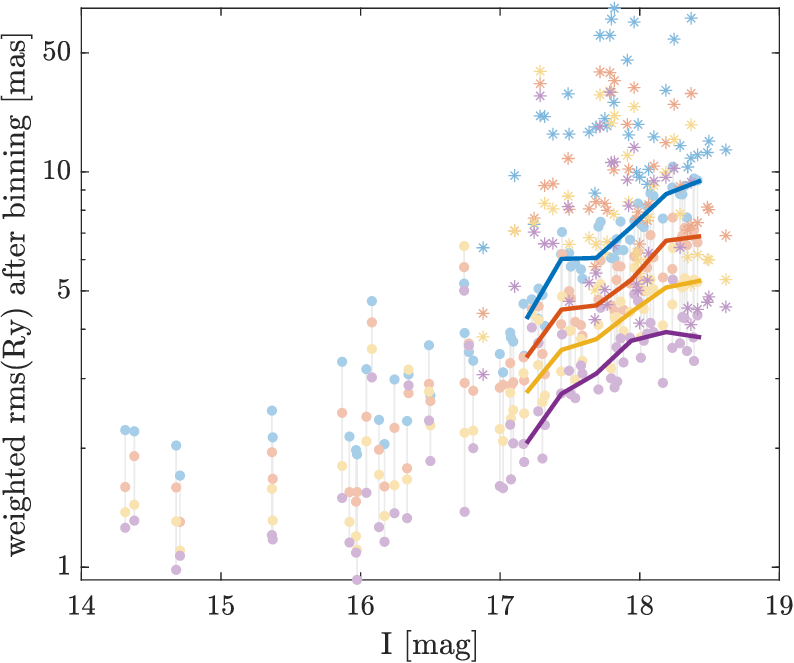}
\caption{Astrometric residual precision versus source magnitude for different time baselines.
For each source, we compute the weighted rms of the residuals about the best-fit astrometric model after binning the residual time series into cadences of 1, 5, 10, and 20 days (colours as indicated). Within each cadence, residuals are averaged per time bin, and the rms of these binned means is calculated with weights proportional to the number of epochs contributing to each bin; only bins containing at least two epochs are retained, and bins with strong outliers are excluded. Thin segments connect measurements of the same source across cadences. Solid lines indicate running median trends in magnitude bins. This short-baseline rms provides an empirical measure of astrometric precision on week--month timescales relevant for astrometric microlensing. Star-level outliers are marked with asterisks.}
\label{fig:rms_vs_I_w_bins}
 \end{figure}

\begin{figure}\centering
\includegraphics[width=1\linewidth]{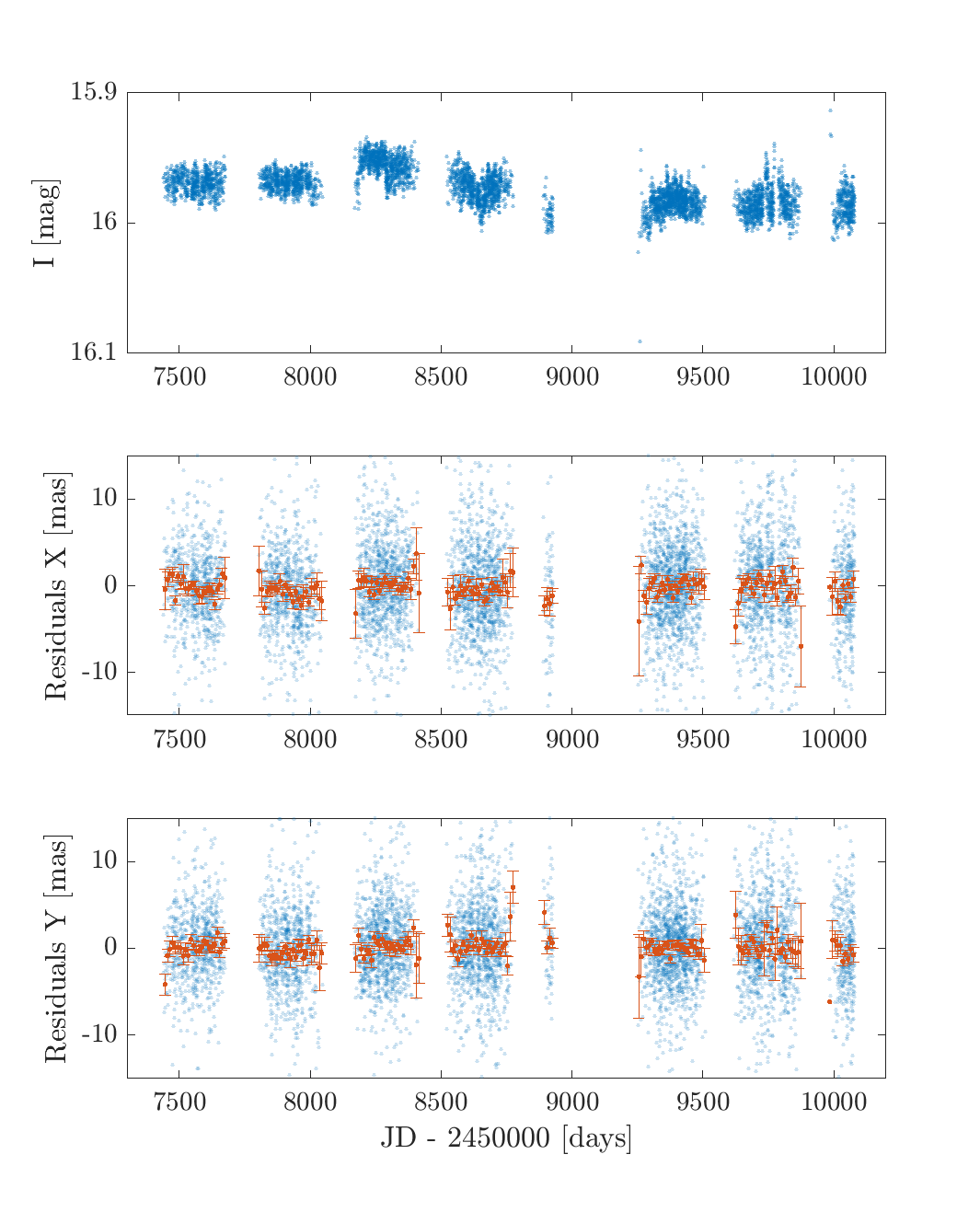}
\caption{An example of the astrometric residuals for a source from KMTNet BLG17K0103. The top panel shows the I-band photometric light curve. The middle and bottom panels display the astrometric residuals in the X and Y directions, respectively. In both astrometric panels, the blue dots are the residuals per epoch, while the orange error bars represent the mean residuals in ten-day bins and their standard error of the mean. The 2D astrometric RMS of the epochal measurements is $\mathord{\sim}$5\,mas, while the RMS of the orange binned points is 1\,mas in both axes.}\label{fig:positioncurve_exmple}
\end{figure}

\begin{figure}\centering
\includegraphics[width=0.9\linewidth]{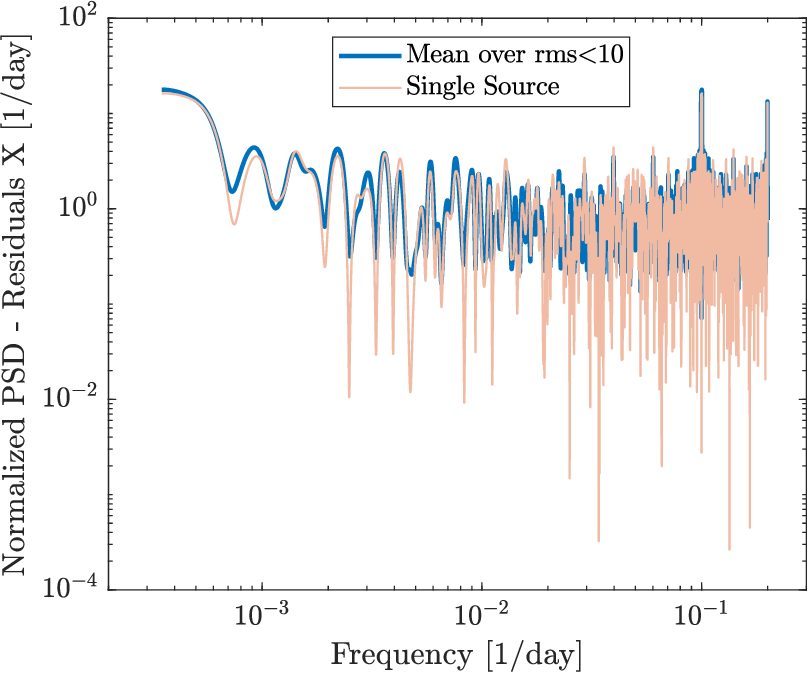}
\includegraphics[width=0.9\linewidth]{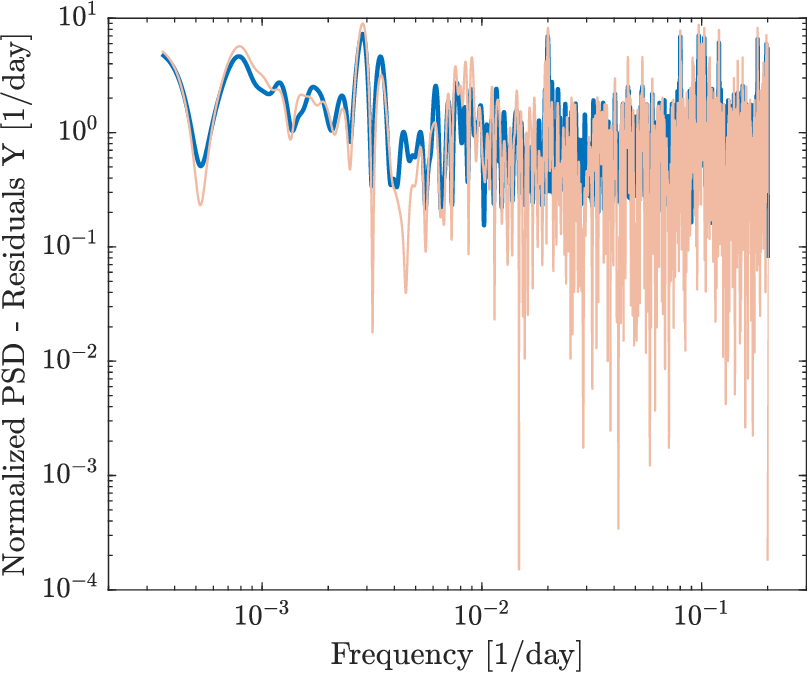}
\caption{Power spectral density (PSD) of the binned position residuals (corresponding to the orange error bars in Figure~\ref{fig:positioncurve_exmple}), shown for the x-axis (top) and y-axis (bottom).
The PSD is normalised by the variance of each time series, making the power dimensionless and enabling comparison between sources.
The blue line indicates the mean PSD over all sources with 2D single epoch precision of $< 10$, while the orange line corresponds to the PSD of the example source shown in Figure~\ref{fig:positioncurve_exmple}.}
\label{fig:positioncurve_PSD_example}
\end{figure}

\section{Test on real data}
\label{sec:results}
We evaluated the performance of our pipeline using images from the KMTNet BLG17K0103 field, taken by the CTIO telescope. The pipeline was applied to image stamps of 300×300 pixels centred around the known microlensing event, KMT-2019-BLG-2630, and limited to sources from the KMTNet catalogue with $I$-band magnitudes of $I\mathord{<}18.5$ whose nearest neighbor (among sources with $I\mathord{<}18.5$) lies at a distance greater than 5\,pixels. This selection yielded 105 selected sources out of 148. Table \ref{tab:ObservationsSum} summarizes the observations for the KMTNet BLG17K0103 field. Observations with an airmass greater than 1.5 or a full width at half maximum (FWHM) exceeding 4.5 pixels (1.8\arcsec) were excluded to minimise the effects of atmospheric distortion and poor seeing conditions. For the PSF modelling (see \S\ref{sec:reduction}), we choose $\bar{r}=$2.5\,pix.

Because our algorithm may be applied in different ways,
here we have used algorithm 6 of Table \ref{tab:experiment_results}. We note that the different algorithmic variations do not result in considerable changes.
Figure \ref{fig:rms_vs_I_w} shows the root-mean-square (rms) of the observed positions of each source around the best-fit astrometric model as a function of $I$-band magnitude.
For sources near the scintillation floor ($13.5 \lesssim I \lesssim 16$), the typical rms per epoch is $\mathord{\sim}3$\,mas in one axis and $\mathord{\sim}5$\,mas in both axes.
Most of the outliers in Figure~\ref{fig:rms_vs_I_w} (marked with 'asterisks') are sources with a nearby companion that do not appear in the KMTNet reference catalogue, or are mistreated as a single PSF. After binning the residual time series in 5--20~day cadences, the bright subset reaches a weighted rms of order $\sim$2~mas per coordinate.
This behaviour is shown in Figure~\ref{fig:rms_vs_I_w_bins} and approaches the precision required for astrometric microlensing.

Figure~\ref{fig:positioncurve_exmple} provides an example of the residuals as a function of time for a single source in the KMTNet BLG17K0103 field. Figure~\ref{fig:positioncurve_PSD_example} shows the corresponding power spectral density (PSD) of the residuals, both for this individual star and for the mean over sources with a 2-D single-epoch precision $<10$\,mas.
In the PSD, there are some indications of red noise presence in the lowest frequencies. This may suggest
that there is room for improvement in the current pipeline model and fitting procedure.

\begin{table*}
\centering
\caption{Summary of the algorithms, detailing the implementation of each pipeline component and the results from the bootstrap and comparison to {\it Gaia}. The comparison includes sources within the top 33$\%$ in single-epoch precision, corresponding to a 2D RMS of $\sim$10\,mas (see Figure~\ref{fig:rms_vs_I_w}). The median reported {\it Gaia} proper motion uncertainty for these sources is 0.12\,mas/yr in right ascension and 0.073\,mas\,year$^{-1}$ in declination. The last two columns list the root-mean-square (rms) of the difference in proper motion between two bootstrap realizations (Boots $\Delta\mu$), and between the pipeline result and {\it Gaia} DR3 ({\it Gaia} $\Delta\mu$). Values are given separately for each coordinate axis. Rows marked with, e.g., \textbf{4×} indicate configurations applied four times.}
\begin{tabular}{c|lllllll|cc}
\toprule
 $n^{int}_{iter}$& \textbf{Algorithm} & \textbf{Weights} & \textbf{PM} & \textbf{DCR} & \textbf{Annual} & \textbf{IntraPixel} & \textbf{SysRem} & \textbf{Boots} rms\,$\Delta \mu$  (x, y)  & \textbf{{\it Gaia}} rms\,$\Delta \mu \, (\alpha,\delta)$ \\
            &   &   &   &   &   &   &   & [mas\,yr$^{-1}$] & [mas\,yr$^{-1}$]\\
\hline\hline
 & \textbf{Algorithm 1}    & \ding{55}& \ding{51}& \ding{55} & \ding{55} & \ding{55} & \ding{55}  &  0.22 , 0.21 &0.40 , 0.24\\
  \cdashline{1-8}
\multirow{1}{*}{\rotatebox[origin=c]{0}{\textbf{4×}}}
 & Step 1 & --& \ding{51}& -- & -- & -- & -- & & \\
\hline
 & \textbf{Algorithm 2}    & \ding{51}& \ding{51}& \ding{55}& \ding{55} & \ding{55} & \ding{51} &  0.12 , 0.11 &0.42 , 0.25 \\
 \cdashline{1-8}
{\rotatebox[origin=c]{0}{\textbf{2×}}}
 & Step 1 & --& \ding{51}& -- & -- & -- & -- & & \\
\cdashline{1-8}
{\rotatebox[origin=c]{0}{\textbf{10×}}}
 & Step 2  & \ding{51}& \ding{51}& --& -- & -- & -- & & \\
 \cdashline{1-8}
 & SysRem & --& --& -- & -- & -- & \ding{51} & & \\
\cdashline{1-8}
{\rotatebox[origin=c]{0}{\textbf{4×}}}
 & Refine   & \ding{51}& \ding{51}& --& -- & -- & -- & & \\
\hline
 & \textbf{Algorithm 3}    & \ding{51}& \ding{51}& \ding{51}& \ding{55} & \ding{55} & \ding{51}  & 0.11 , 0.098 &0.41 , 0.25\\
\cdashline{1-8}
{\rotatebox[origin=c]{0}{\textbf{2×}}}
 & Step 1 & --& \ding{51}& -- & -- & -- & -- & & \\
 \cdashline{1-8}
{\rotatebox[origin=c]{0}{\textbf{10×}}}
 & Step 2   & \ding{51}& \ding{51}& \ding{51}& -- & -- & -- & & \\
 \cdashline{1-8}
 & SysRem & --& --& -- & -- & -- & \ding{51} & & \\
 \cdashline{1-8}
{\rotatebox[origin=c]{0}{\textbf{4×}}}
 & Refine & \ding{51}& \ding{51}& \ding{51}& -- & -- & -- & & \\
\hline
 & \textbf{Algorithm 4}    & \ding{51}& \ding{51}& \ding{55}& \ding{51} & \ding{55} & \ding{51} & 0.12 , 0.11  &0.41 , 0.25 \\
\cdashline{1-8}
{\rotatebox[origin=c]{0}{\textbf{2×}}}
 & Step 1 & --& \ding{51}& -- & -- & -- & -- & & \\
 \cdashline{1-8}
{\rotatebox[origin=c]{0}{\textbf{10×}}}
 & Step 2  & \ding{51}& \ding{51}& --& \ding{51} & -- & -- & & \\
 & Annual effect  & --& --& --& \ding{51} & -- & -- & & \\
 \cdashline{1-8}
 & SysRem & --& --& -- & -- & -- & \ding{51} & & \\
 \cdashline{1-8}
{\rotatebox[origin=c]{0}{\textbf{4×}}}
 & Refine   & \ding{51}& \ding{51}& --& \ding{51} & -- & -- & & \\
\hline
 & \textbf{Algorithm 5}    & \ding{51}& \ding{51}& \ding{51}& \ding{51} & \ding{55} & \ding{51} & 0.10 , 0.091 & 0.40 , 0.24 \\
\cdashline{1-8}
{\rotatebox[origin=c]{0}{\textbf{2×}}}
 & Step 1 & --& \ding{51}& -- & -- & -- & -- & & \\
 \cdashline{1-8}
{\rotatebox[origin=c]{0}{\textbf{10×}}}
 & Step 2   & \ding{51}& \ding{51}& \ding{51}& -- & -- & -- & & \\
 & Annual effect  & --& --& --& \ding{51} & -- & -- & & \\
 \cdashline{1-8}
 & SysRem & --& --& -- & -- & -- & \ding{51} & & \\
 \cdashline{1-8}
{\rotatebox[origin=c]{0}{\textbf{4×}}}
 & Refine   & \ding{51}& \ding{51}& \ding{51}& -- & -- & -- & & \\
\hline
 & \textbf{Algorithm 6}    & \ding{51}& \ding{51}& \ding{51}& \ding{51} & \ding{51} & \ding{51} & 0.10 , 0.091  &0.39 , 0.23 \\
 \cdashline{1-8}
{\rotatebox[origin=c]{0}{\textbf{2×}}}
 & Step 1 & --& \ding{51}& -- & -- & -- & -- & & \\
 \cdashline{1-8}
{\rotatebox[origin=c]{0}{\textbf{10×}}}
 & Step 2  & \ding{51}& \ding{51}& \ding{51}& -- & -- & -- & & \\
 & Annual effect  & --& --& --& \ding{51} & -- & -- & & \\
 & IntraPixel  & --& --& --& -- & \ding{51} & -- & & \\
 \cdashline{1-8}
 & SysRem & --& --& -- & -- & -- & \ding{51} & & \\
 \cdashline{1-8}
{\rotatebox[origin=c]{0}{\textbf{4×}}}
 & Refine   & \ding{51}& \ding{51}& \ding{51}& -- & -- & -- & & \\
\bottomrule
\end{tabular}
\label{tab:experiment_results}
\end{table*}

\subsection{Bootstrap}
\label{sec:bootstrap_comparison}

To estimate the precision and consistency of our results, we perform a Bootstrap test (\citealt{efron1982jackknife}). We split the dataset into two subsets (i.e., even and odd indices), run the pipeline separately for each subset, and compare the results. The Bootstrap test is conducted for multiple scenarios, each incorporating different components of the astrometric model.

\begin{figure*}
    \subfloat[]{%
        \includegraphics[width=.48\linewidth]{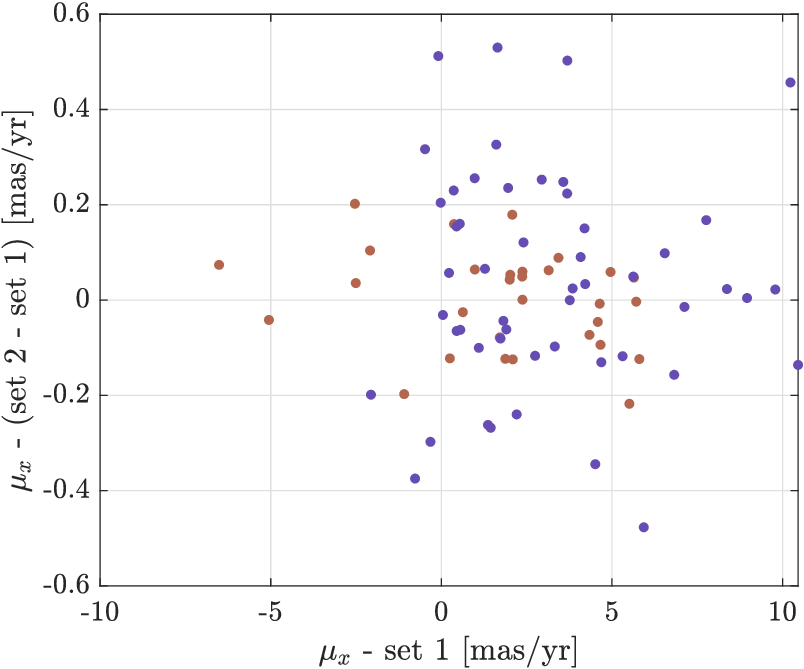}%
        \label{subfig:a}%
    }\hfill
    \subfloat[]{%
        \includegraphics[width=.48\linewidth]{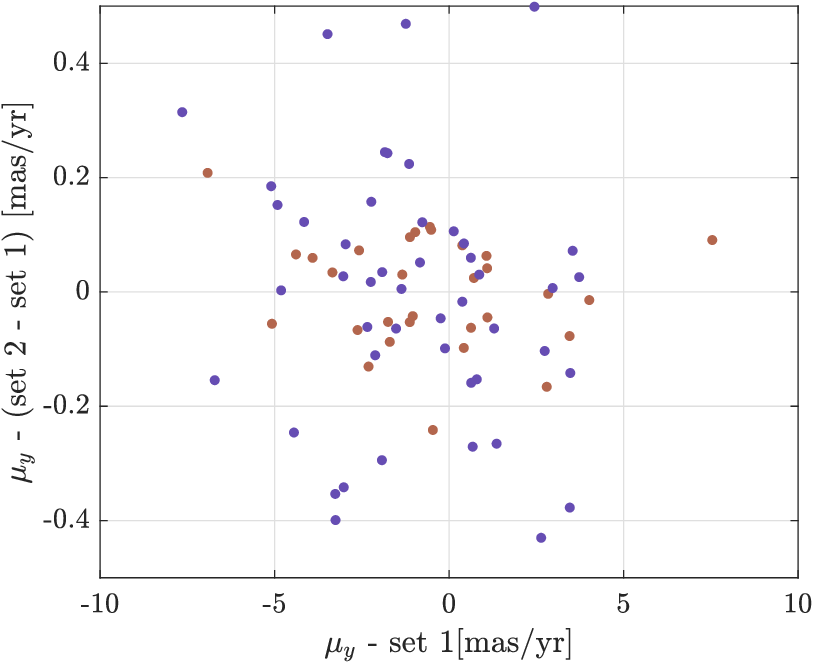}%
         \label{subfig:b}%
    }\\
    \subfloat[]{%
        \includegraphics[width=.48\linewidth]{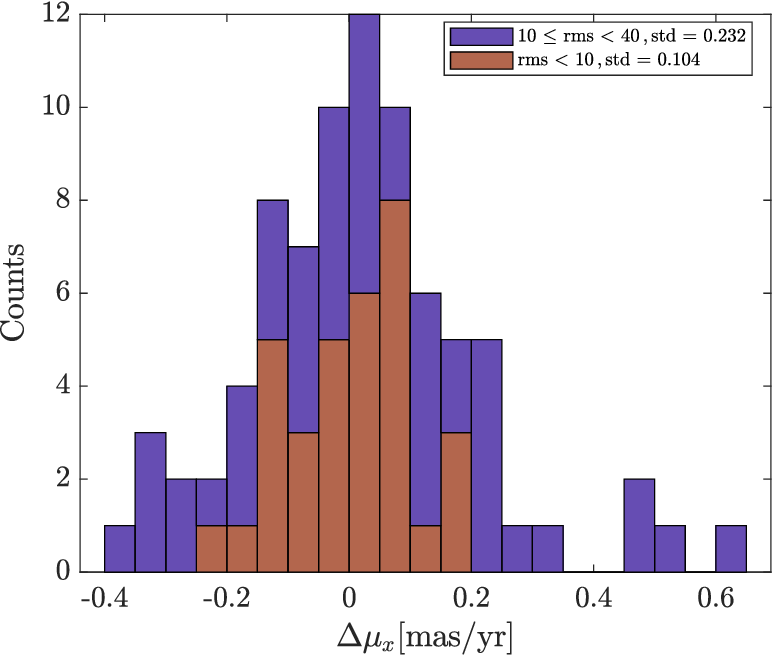}%
        \label{subfig:c}%
    }\hfill
    \subfloat[]{%
        \includegraphics[width=.48\linewidth]{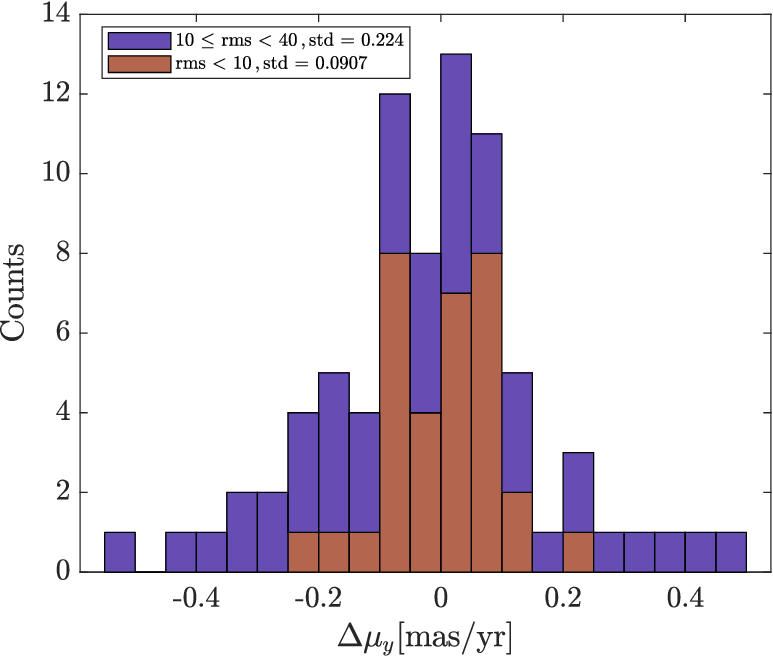}%
        \label{subfig:d}%
    }
    \caption{Bootstrap test for proper motion in the BLG17K0103 field.  Figure~\ref{subfig:a} (Figure~\ref{subfig:b}) shows the difference in proper motion along the x-axis (y-axis) between the two datasets, plotted as a function of the proper motion from dataset 1.  Figure \ref{subfig:c} (\ref{subfig:d}) shows the histogram of the x-axis (y-axis) proper motion difference between the two data sets. The violet colour denotes sources with single-epoch precision between 10 and 40\,mas, while the red colour indicates sources with single-epoch precision better than 10\,mas.} 
    \label{fig:twosets_comparison_17K0103}
\end{figure*}

\begin{figure}\centering
\includegraphics[width=1\linewidth]{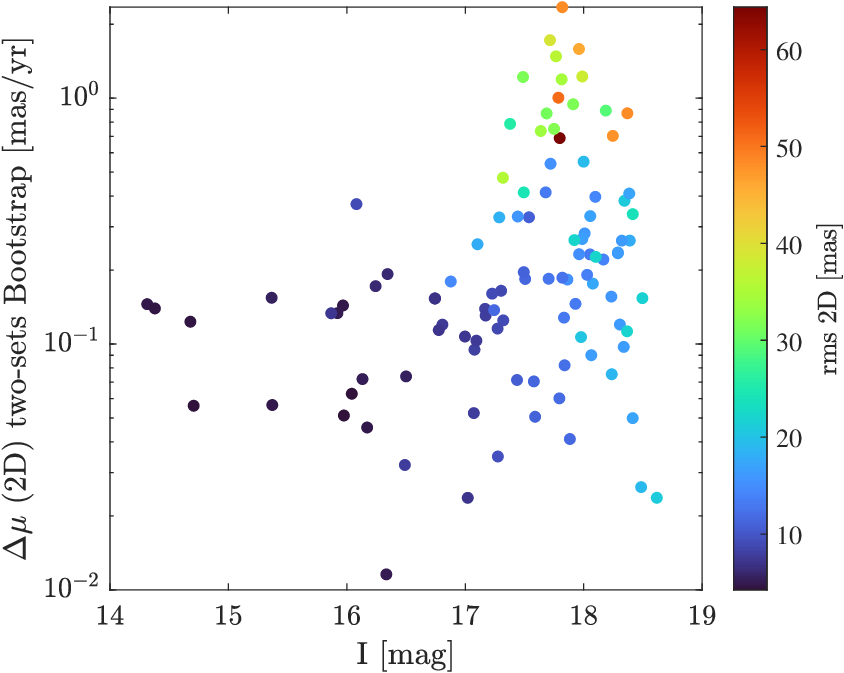}
\caption{2D proper motion difference between the two data sets of field BLG17K0103, used in the Bootstrap test, shown as a function of source magnitude. The 2D difference is calculated as the square root of the sum of squares of the proper motion components along each axis. The colorbar indicates the 2D precision of each source, computed as the root-mean-square (rms) of the residuals from the best-fit model.}
\label{fig:Bootstrap_pm2D_vs_mag}
\end{figure}

Table~\ref{tab:experiment_results} provides the Bootstrap rms for different algorithm configurations.
Figure~\ref{fig:twosets_comparison_17K0103} compares the proper motion measurements between the two subsets for Algorithm 6 (see Table~\ref{tab:experiment_results}), and Figure~\ref{fig:Bootstrap_pm2D_vs_mag} shows the 2D proper motion differences as a function of magnitude. The comparison is shown for two groups of sources: (1) sources with a single-epoch precision of $<10$\,mas (orange) and (2) sources with a single-epoch precision between $10$ and $40$\,mas (blue). Additionally, the comparison indicates a precision level of approximately\footnote{Reduced to $0.07$\,mas\,year$^{-1}$ if one considers that the number of observations were reduced by a factor of two in the Bootstrap test.} $0.1$\,mas\,year$^{-1}$. The best theoretical precision expected for this dataset is about $0.04$\,mas ($\approx 10/\sqrt{3200}$), where 10\,mas represents the per-epoch precision, and 3200 is the assumed number of epochs. Consequently, the precision inferred from the bootstrap analysis is roughly a factor of 2--3 worse than the theoretical expectation.

\subsection{Comparison with Gaia DR3}
\label{sec:compare_gaia}

\begin{figure}
\includegraphics[width=0.95\linewidth]{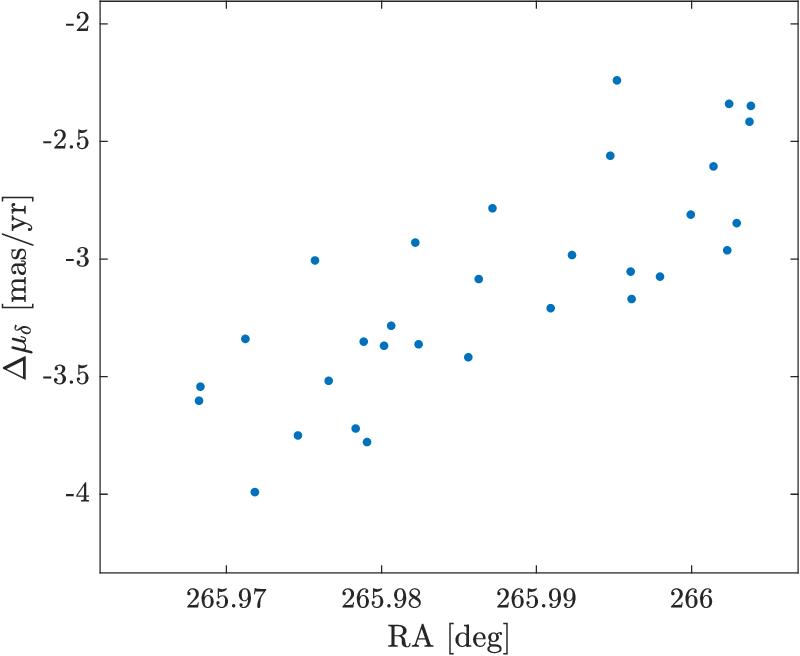}
\caption{Example of proper motion comparison between KMTNet BLG17K0103 and Gaia DR3 before applying Equations~\ref{eq:GaiaAffineTranAlpha} and~\ref{eq:GaiaAffineTranDelta}. Here, proper motion in declination vs. the source's right ascension in Gaia DR3. }
\label{fig:gaia_comp_position_corr}
\end{figure}

\begin{figure*}
    \subfloat[]{%
        \includegraphics[width=.48\linewidth]{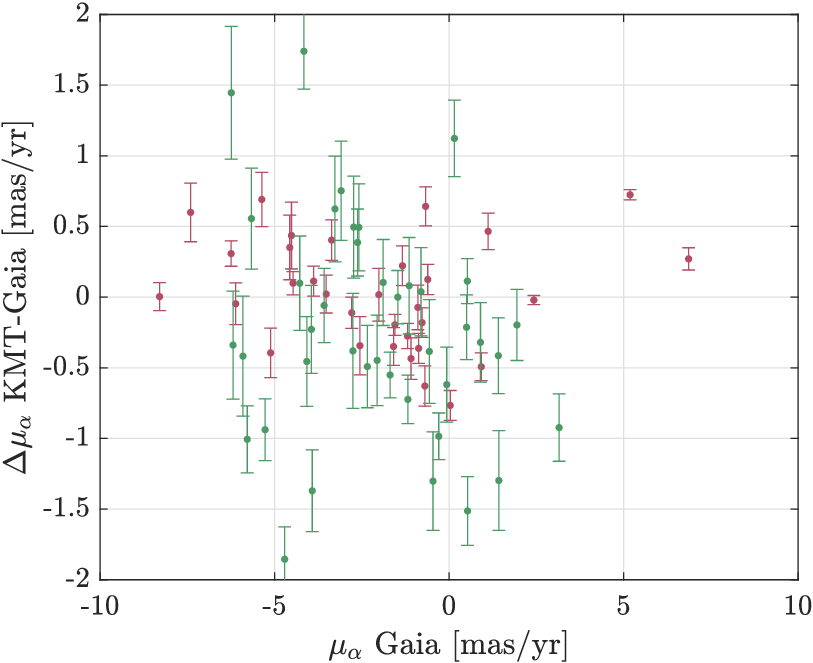}%
        \label{subfig:gaia_comp_a}%
    }\hfill
    \subfloat[]{%
        \includegraphics[width=.48\linewidth]{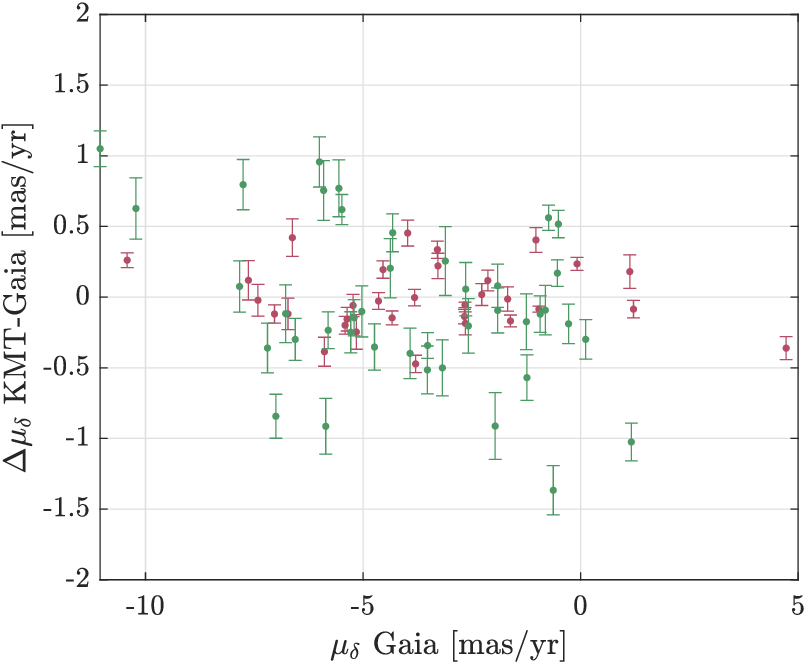}%
        \label{subfig:gaia_comp_b}%
    }\\
    \subfloat[]{%
        \includegraphics[width=.48\linewidth]{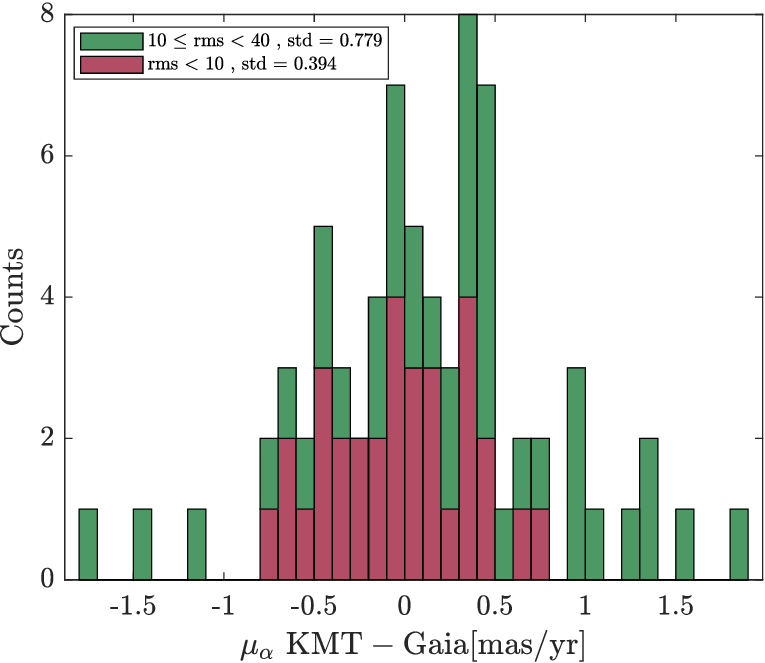}%
        \label{subfig:gaia_comp_c}%
    }\hfill
    \subfloat[]{%
        \includegraphics[width=.48\linewidth]{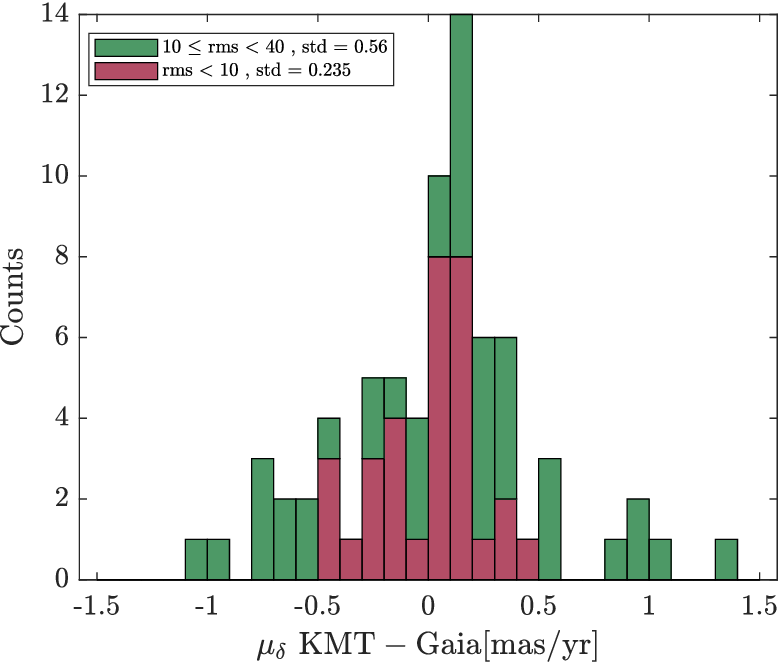}%
        \label{subfig:gaia_comp_d}%
    }
    \caption{Comparison of proper motions in the 17K0103 field with {\it Gaia} DR3. Figures \ref{subfig:gaia_comp_a} and \ref{subfig:gaia_comp_b} show proper motion differences in right ascension and declination (KMTNet vs. {\it Gaia} DR3), respectively, as a function of the {\it Gaia} DR3 proper motion. Figures \ref{subfig:gaia_comp_c} and \ref{subfig:gaia_comp_d} display histograms of proper motion differences in right ascension and declination after outlier removal. The legend indicates the standard deviation of proper motion differences under specific conditions. The bin size is 0.1\,mas\,yr$^{-1}$.}    
    \label{fig:gaia_comp}
\end{figure*}

\begin{figure}
\includegraphics[width=1\linewidth]{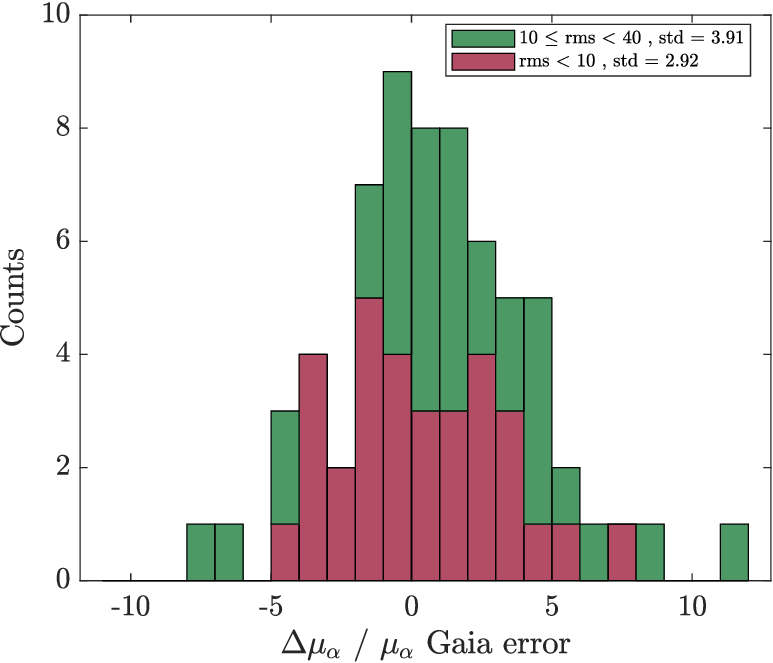}
\includegraphics[width=1\linewidth]{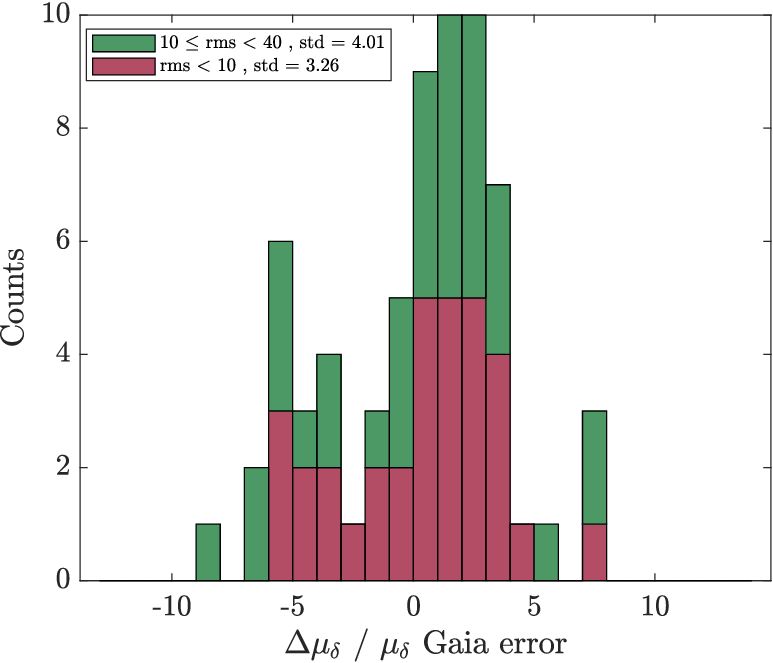}
\caption{The ratio of the proper motion difference to the Gaia DR3 error in KMTNet BLG17K0103. Top: right ascension, bottom: declination. The legend indicates the standard deviation of proper motion differences under specific conditions. The bin size is 1.}
\label{fig:gaia_comp_chi2}
\end{figure}

Due to the high density of stars in the Galactic bulge, {\it Gaia} measurements in this region are sparse and even missing for many stars.
Nevertheless, we attempted to compare our results with {\it Gaia}.
We cross-matched our sources with {\it Gaia} DR3 \citep{GAIA_mission_prusti2016gaia, GAIA+2022yCat_GAIA_DR3_MainSourcesCatalog}, excluding outliers (28 out of 105 sources; see \S\ref{sec:weights}) as well as sources with {\tt ruwe}$>$1.3 (5 additional sources, of which two were already among the outliers).
Figure \ref{fig:gaia_comp_position_corr} shows the differences between the {\it Gaia} and our measured proper motion as a function of the stellar right ascension and declination.
Although there is a clear linear relation between the {\it Gaia} and KMTNet-based measurements, as expected, these relations are not one-to-one, and they show a clear offset.
The reason for this is that while {\it Gaia} is measuring proper motion in a global coordinate system (ICRS),
while the KMTNet measurements are done with respect to a relative coordinate system.
Furthermore, while the mean stellar environment (and proper motion) may change along the KMTNet fields, this may introduce
some complicated relations between the {\it Gaia} and KMTNet proper motion systems.
To take this into account, we fitted the difference in proper motion to
\begin{eqnarray}    
   \mu_{\alpha, {\rm GAIA}} = \Delta{\mu_{\alpha, {\rm KMT}}} + s_{\alpha,\alpha}\alpha + s_{\alpha,\delta}\delta + \label{eq:GaiaAffineTranAlpha}\\ q_{\alpha,\alpha}\mu_{\alpha, {\rm KMT}} + q_{\alpha,\delta}\mu_{\delta, {\rm KMT}},\nonumber\\
   \mu_{\delta, {\rm GAIA}} = \Delta{\mu_{\delta, {\rm KMT}}} + s_{\delta,\alpha}\alpha + s_{\delta,\delta}\delta +\label{eq:GaiaAffineTranDelta}\\ \nonumber q_{\delta,\alpha}\mu_{\alpha, {\rm KMT}} + q_{\delta,\delta}\mu_{\delta, {\rm KMT}},
\end{eqnarray}
The values of the parameters are field-dependent.
Figure~\ref{fig:gaia_comp} compares the {\it Gaia} and KMTNet proper motions after this fit was performed.
In Table~\ref{tab:experiment_results}, we also provide the rms compared to {\it Gaia} proper motion measurements after fitting this transformation.
We note that the median of the {\it Gaia} proper motion errors (accuracy) for the sources we use for the comparison is 0.073\,mas\,year$^{-1}$ (0.12\,mas\,year$^{-1}$) for declination (right ascension),  which are comparable with our proper motion precision, as can be seen in Figure \ref{fig:gaia_comp_chi2} which shows the ratio between the difference of proper motion ({\it Gaia}-KMT) and the {\it Gaia} proper motion accuracy. The observed error ratio, approximately $\mathord{\sim}3$, may also be indicative of an underestimation of the {\it Gaia} uncertainties in the Galactic Bulge.
This underestimation of the GAIA errors in the bulge is also supported by  \citet{vasiliev2021gaia_error_underestimation} and \citet{luna2023Gaia_astrometry_underestimation}.

\subsubsection{Comparison of Absolute Positions with {\it Gaia}}
\label{sec:compare_gaia_positions}

In addition to the proper-motion comparison described above, we also examined the consistency of our measured absolute positions with those from {\it Gaia} DR3. The comparison was performed in the same manner as the proper-motion analysis, using the affine transformations defined in Equations~\ref{eq:GaiaAffineTranAlpha} and~\ref{eq:GaiaAffineTranDelta}. 

\begin{figure}
\includegraphics[width=1\linewidth]{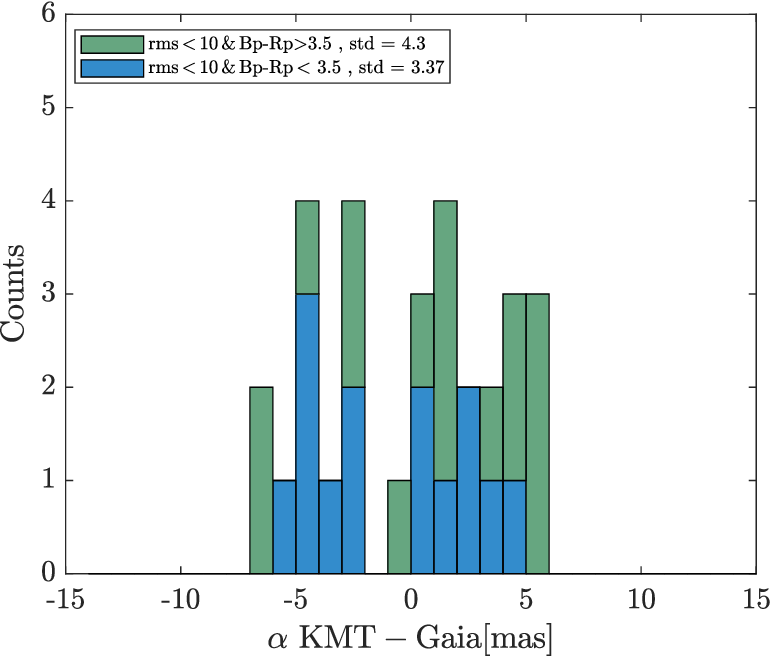}
\includegraphics[width=1\linewidth]{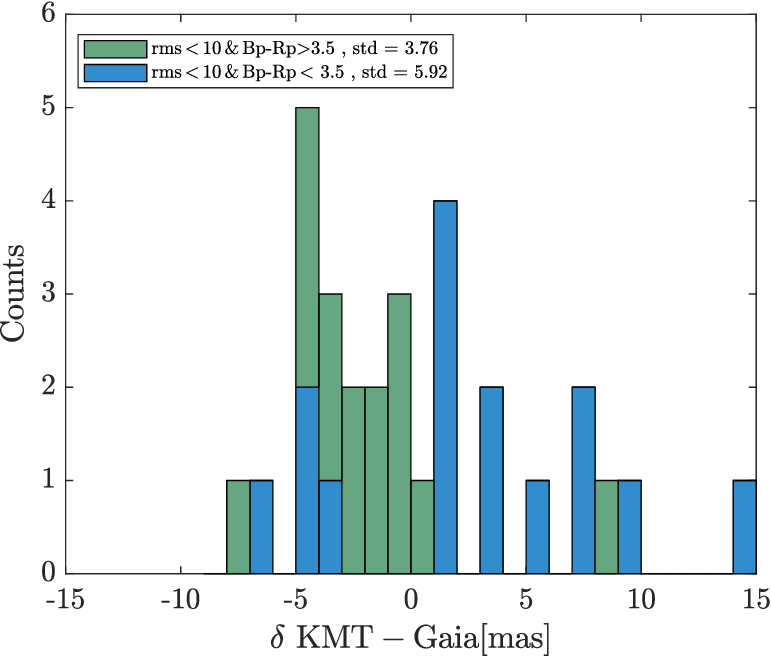}
\caption{Comparison of absolute positions in the 17K0103 field with {\it Gaia} DR3. The upper and lower panels show histograms of positional differences in right ascension and declination (KMTNet~--~{\it Gaia}), respectively. Sources are divided by colour, with blue and green histograms representing stars with ${\rm Bp-Rp}<3.5$ and ${\rm Bp-Rp}\ge3.5$, respectively. The legend indicates the standard deviation of the positional differences for each subsample. Only sources with rms$<10$\,mas are included. The bin size is 1\,mas.}
\label{fig:gaia_comp_position_color}
\end{figure}

A direct comparison between the absolute positions derived from our relative astrometric solution and those from {\it Gaia} DR3, shown in Figure~\ref{fig:gaia_comp_position_color}, reveals that the positional scatter (rms $\approx4$\,mas) is roughly twice as large as expected from the formal uncertainties ($\approx1.5$\,mas, estimated by multiplying the proper-motion rms by the 7\,yr time baseline). Such an excess is naturally expected for a relative astrometric solution, since the global reference frame is free to drift with respect to the inertial system. The excess scatter likely originates from residual systematics inherent to the relative nature of our solution. Because the astrometry is solved relative to the mean reference frame rather than the ICRS, slow variations in the effective zero point of the frame, driven by seasonal atmospheric conditions or colour-dependent refraction, can shift the local coordinate system with respect to {\it Gaia}. In particular, the annual effect (§\ref{sec:annual_effects_plx}), which reflects residual colour- and season-dependent refraction terms not fully captured by the basic chromatic model, may introduce correlated positional offsets that vary on yearly timescales. Such offsets, combined with small residual geometric distortions or long-term drift of the fiducial frame, can naturally lead to the observed excess scatter.

To test whether colour-dependent effects contribute to the observed scatter, we divided the sources into two colour bins, ${\rm Bp-Rp}<3.5$ and ${\rm Bp-Rp}\ge3.5$, and repeated the position comparison for each subset (Figure~\ref{fig:gaia_comp_position_color}). The resulting distributions show comparable dispersions, with only a modest difference between the blue and green subsamples, suggesting that while chromatic systematics may be present, they are not the dominant source of the observed positional scatter.

It is important to note that even systematics that are partially absorbed by the affine or chromatic terms can propagate into the reference positions $(x_{0},y_{0})$, since these parameters are derived from the same global fit. Consequently, any residual frame-dependent offsets, whether seasonal, chromatic, or geometric, manifest as excess scatter in the absolute position comparison with {\it Gaia}, while having a much smaller impact on the relative proper motions. Therefore, the proper-motion comparison with {\it Gaia} remains reliable, although it should be interpreted as relative rather than tied to an absolute reference frame.

\subsection{Cross-Observatory Consistency: CTIO vs. SAAO on BLG15M0306}

We compared the astrometric solutions obtained independently from CTIO and SAAO observations of the KMTNet BLG15M0306 field. 
We selected BLG15M0306 because the SAAO images of KMTNet BLG17K0103 suffer from a significant number of bad columns, whereas BLG15M0306 observed from SAAO exhibits similar observational characteristics to BLG17K0103 (see Table~\ref{tab:ObservationsSum}).

The same selection criteria and quality cuts described earlier for BLG17K0103 were applied, including limiting the sample to sources brighter than $I\mathord{<}18.5$, with nearest-neighbour distances greater than 5\,pixels, and excluding observations with an airmass $>1.5$ or FWHM $>4.5$ pixels (1.8\arcsec).

\begin{figure*}
    \subfloat[]{%
        \includegraphics[width=.48\linewidth]{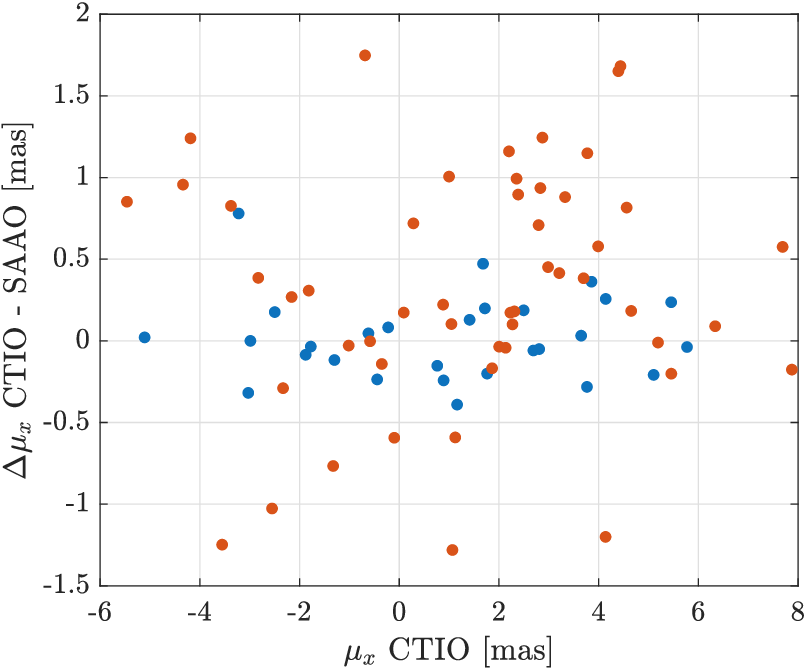}%
        \label{subfig:twotel_a}%
    }\hfill
    \subfloat[]{%
        \includegraphics[width=.48\linewidth]{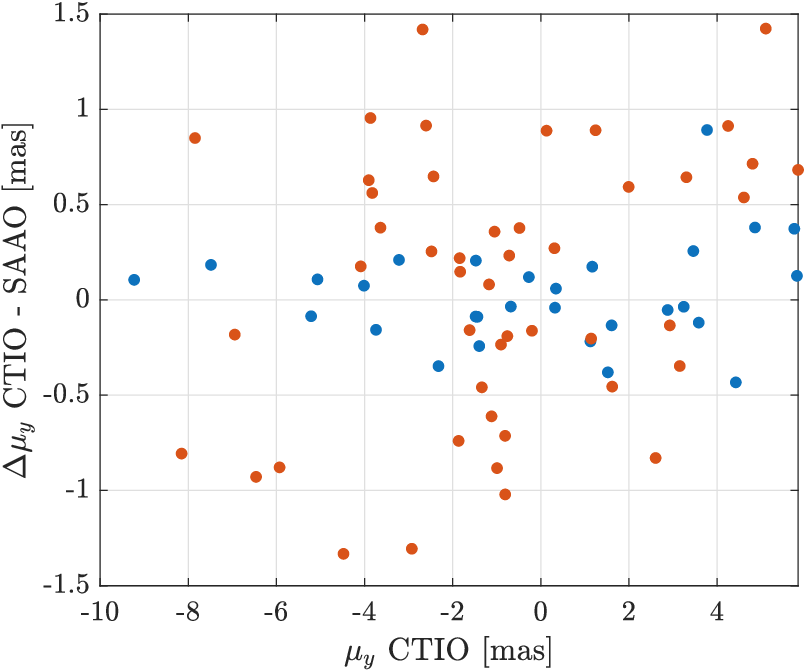}%
         \label{subfig:twotel_b}%
    }\\
    \subfloat[]{%
        \includegraphics[width=.48\linewidth]{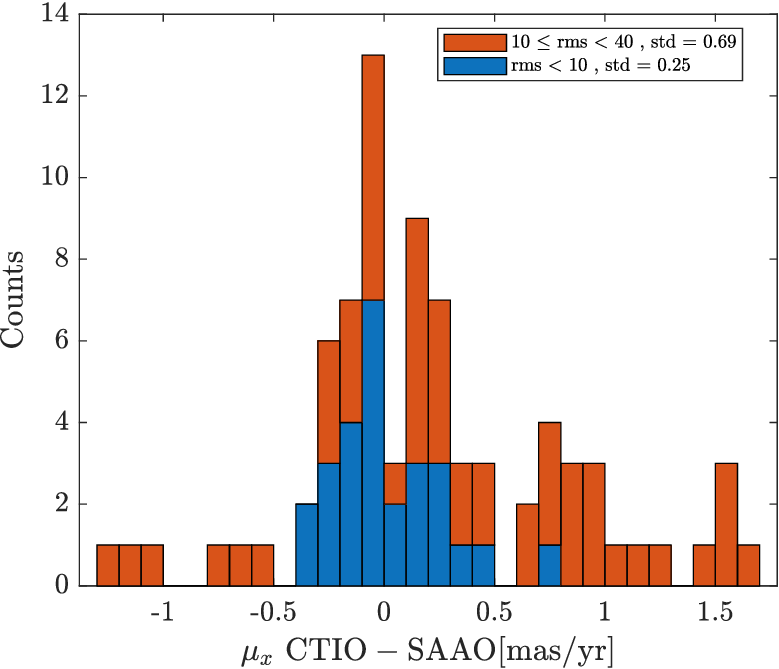}%
        \label{subfig:twotel_c}%
    }\hfill
    \subfloat[]{%
        \includegraphics[width=.48\linewidth]{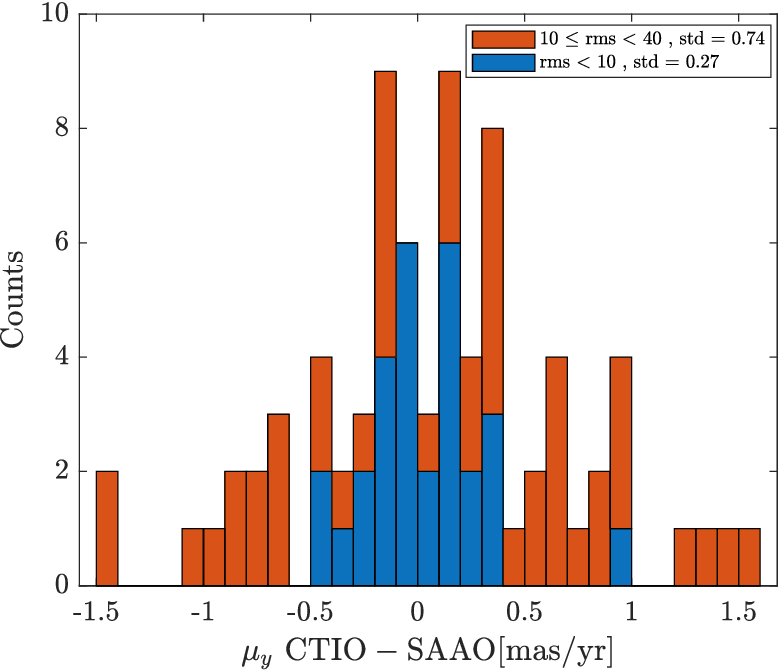}%
        \label{subfig:twotel_d}%
    }
    \caption{Comparison of proper motions, between CTIO and SAAO observatories, in the BLG15M0306 field. Figure~\ref{subfig:twotel_a} (\ref{subfig:twotel_b}) shows the x-axis (y-axis) proper motion difference between the two telescopes, CTIO-SAAO, as a function of the proper motion measured in CTIO. Figure~\ref{subfig:twotel_c} (\ref{subfig:twotel_d}) shows the histogram of the x-axis (y-axis) proper motion differences between the two observatories. The blue colour represents sources with single-epoch precision better than 10\,mas, while the red colour represents sources with precision between 10 and 40\,mas.} 
    \label{fig:TelescopeComparison_15M03063}
\end{figure*}

\begin{figure}\centering
\includegraphics[width=1\linewidth]{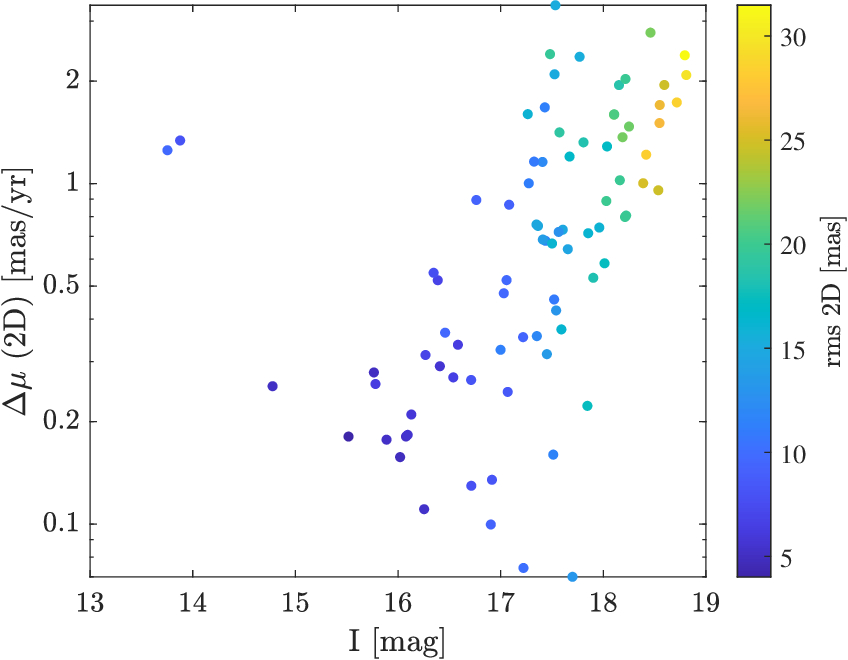}
\caption{2D proper motion difference between CTIO and SAAO observatories in the BLG15M0306 field, shown as a function of source magnitude. The 2D difference is calculated as the square root of the sum of squares of the proper motion components along each axis. The colorbar indicates the single-epoch 2D precision of each source, computed as the root-mean-square (rms) of the residuals from the best-fit model.}
\label{fig:TwoTelescopes_pm_vs_mag}
\end{figure}

For each star, we extracted the fitted parameters \((x_0, y_0, \mu_x, \mu_y)\) from both datasets. 
The differences in proper motions were modelled as linear functions of position and motion, following the form of Equations~\ref{eq:GaiaAffineTranAlpha} and~\ref{eq:GaiaAffineTranDelta}.
After applying these corrections, the RMS scatter of the proper motion differences was $\mathord{\sim}0.25$\,mas\,year$^{-1}$.
The precision estimated from the bootstrap analysis for this field is approximately $0.1$\,mas\,yr$^{-1}$ for the CTIO data and $0.15$\,mas\,yr$^{-1}$ for the SAAO observations. Based on these values, the expected precision for the two-telescope comparison is $\sim$1.8\,mas\,yr$^{-1}$. This is compatible with, though slightly higher than, the precision actually obtained in the comparison. The discrepancy may indicate a slight underestimation of the bootstrap uncertainties or suggest a small systematic offset between the two telescopes.
No large-scale trends remained in the corrected residuals.
The comparison is illustrated in Figure~\ref{fig:TelescopeComparison_15M03063}, which shows the difference between measured proper motions from CTIO and SAAO, as well as the distribution of their differences after correction. Figure~\ref{fig:TwoTelescopes_pm_vs_mag} further presents two-dimensional proper motion difference as a function of source magnitude.
These results demonstrate the reproducibility and cross-consistency of the astrometric solutions across independent observatories within our pipeline.

\section{Conclusion}
\label{sec:conclusion}
We presented a pipeline for precise astrometric measurements of ground-based observations, aiming to achieve milliarcsecond-level astrometric precision necessary for detecting isolated compact objects via astrometric microlensing.
The pipeline is available online\footnote{\textit{Code availability:} The Lizoo pipeline and installation instructions are available at \url{https://github.com/noamse/Lizoo}.}

The pipeline was demonstrated on approximately $\mathord{\sim}6500$ images from the KMTNet BLG17K0103 field, successfully measuring and correcting for systematic effects such as Differential Chromatic Refraction (DCR) and the Annual effect (see \S\ref{sec:chromatic_correction} and \S\ref {sec:annual_effects_plx}). At the bright end (mag $\lesssim$17), we reach a per-epoch astrometric precision of $\mathord{\sim}$5\,mas.
In addition, we quantified short-cadence astrometric precision by binning
the residual time series over 5--20~day cadences.
For bright sources, this stacking reduces the weighted rms to
$\sim$2~mas per coordinate (Figure~\ref{fig:rms_vs_I_w_bins}),
close to the precision required for astrometric microlensing, and the
consistent improvement across cadences indicates that the residual scatter scales down with temporal stacking. 

To evaluate the precision of the proper motions derived, we use the Bootstrap technique and compare the results with Gaia DR3 data. The Bootstrap test indicates a precision of $\mathord{\sim}0.1$ mas/year, while the comparison with Gaia DR3 yields proper motion precisions of $\mathord{\sim}0.3$\,mas\,year$^{-1}$. The precision as estimated compared to Gaia DR3, is a factor of $\mathord{\sim}3$ higher than the precision estimated from the Bootstrap test.
This may suggest that either the precision of the Gaia-DR3 proper motion measurements in the bulge is underestimated
(which is supported by \citealt{vasiliev2021gaia_error_underestimation,
luna2023Gaia_astrometry_underestimation}), and/or there is an additional systematic error in our measurements.

Because astrometric microlensing is a relative astrometry phenomenon, certain systematics may have minimal impact or can be corrected. The dependence of precision on source position, as shown in Figure \ref{fig:gaia_comp_position_corr}, underscores the importance of addressing these factors.

Our comparison of the astrometric solutions from CTIO and SAAO demonstrates that combining data from multiple telescopes has the potential to further improve astrometric precision and robustness. 
The independent datasets exhibit consistent solutions after appropriate calibration, suggesting that a joint fit leveraging both observatories could reduce random errors and enhance temporal sampling.
However, such a combination must be approached carefully, accounting for field-dependent systematics, instrument-specific biases, and potential distortions introduced by differing atmospheric and instrumental conditions.

Although we successfully measured and corrected some systematics, the achieved precision remains approximately 2--3 times worse than the theoretical Poisson noise limit. This indicates the presence of additional undetected systematics that were not modelled.

The KMTNet dataset, with its high cadence and extensive coverage of microlensing events, provides a unique opportunity to search for isolated black holes and constrain their mass function. Future work will focus on measuring proper motion in the KMTNet fields and searching for isolated Black Holes.

\section*{Acknowledgements}

This research has made use of the KMTNet system
operated by the Korea Astronomy and Space Science Institute
(KASI) at three host sites of CTIO in Chile, SAAO in South
Africa, and SSO in Australia. Data transfer from the host site to
KASI was supported by the Korea Research Environment
Open NETwork (KREONET). This research was supported by KASI
under the R\&D program (project No. 2024-1-832-01) supervised
by the Ministry of Science and ICT.


E.O.O. is grateful for the support of
grants from the 
Willner Family Leadership Institute,
André Deloro Institute,
Paul and Tina Gardner,
The Norman E Alexander Family M Foundation ULTRASAT Data Center Fund,
Israel Science Foundation,
Israeli Ministry of Science,
Minerva,
BSF, BSF-transformative, NSF-BSF,
Israel Council for Higher Education (VATAT),
Sagol Weizmann-MIT,
Yeda-Sela, and the
Rosa and Emilio Segr\`e Research Award.
This research is supported by the Israeli Council for Higher Education (CHE) via the Weizmann Data Science Research Center, and by a research grant from the Estate of Harry Schutzman.

W.Zang, H.Y., S.M., R.K., J.Z., and W.Zhu acknowledge support by the National Natural Science Foundation of China (Grant No. 12133005).

W.Zang acknowledges the support from the Harvard-Smithsonian Center for Astrophysics through the CfA Fellowship.

J.C.Y. and I.-G.S. acknowledge support from U.S. NSF Grant No. AST-2108414.

Work by C.H. was supported by the grants of National Research Foundation of Korea (2019R1A2C2085965 and 2020R1A4A2002885).

J.C.Y. acknowledges support from a Scholarly Studies grant from the Smithsonian Institution.

\section*{Data Availability}

The raw KMTNet images used in this work are not publicly available but can be accessed upon request through the Korea Astronomy and Space Science Institute (KASI) and subject to their data-sharing policies. Derived data products and astrometric catalogues generated by our pipeline will be shared on reasonable request to the corresponding author.
The Lizoo astrometric pipeline code is available on GitHub at \url{https://github.com/noamse/Lizoo}.
\bibliographystyle{mnras}
\bibliography{papers,HPastrometry}
\appendix
\begin{appendices}

\section{PSF - Multivariate t-distribution}
\label{sec:mtd}

To mitigate noise in the wings of the PSF, such as negative values, we fit the PSF's wings with a continuous function. 

The Multivariate t-Distribution (MTD), a generalization of the Moffat function, is well-suited for PSF fitting, as light scattering significantly affects the PSF wings, which can exhibit strong asymmetry. 

The probability density function (PDF) of the MTD is given by:

\begin{align}
    f(\textbf{x} \mid \nu, p, \mu, \Sigma) &= \frac{\Gamma\left(\frac{\nu + p}{2}\right)}{\Gamma\left(\frac{\nu}{2}\right) \nu^{p/2} \pi^{p/2} \left|\Sigma\right|^{1/2}} \\
    &\quad \times
    \left[1 + \nu^{-1} \left(\textbf{x} - \mu\right)^T \Sigma^{-1} \left(\textbf{x} - \mu\right) \right]^{-(\nu + p)/2}
\end{align}
where $\nu$ is the degrees of freedom, $p$ is the dimension of $\textbf{x}$ (2 in our case), and $\Sigma$ is a $2\times2$ covariance matrix.

\end{appendices}
\end{document}